\documentclass{vgtc}
\usepackage{cite}
\usepackage{amsmath,amssymb,amsfonts}
\usepackage{algorithm}
\usepackage[noend]{algpseudocode}
\usepackage{graphicx}
\usepackage{caption}
\usepackage{textcomp}
\usepackage[colorlinks=true,linkcolor=blue]{hyperref}%
\algblock{ParFor}{EndParFor}
\algnewcommand\algorithmicparfor{\textbf{for}}
\algnewcommand\algorithmicpardo{\textbf{do in parallel}}
\algnewcommand\algorithmicendparfor{}
\algrenewtext{ParFor}[1]{\algorithmicparfor\ #1\ \algorithmicpardo}
\algrenewtext{EndParFor}{\textbf{end parallel for}}

\makeatother
\usepackage{float}
\usepackage{mathtools} 
\usepackage{mathtools}
\usepackage{graphics}
\usepackage{placeins}
\usepackage{hyperref}
\usepackage{multirow}
\usepackage{array}
\usepackage{enumitem}
 \usepackage{color}
 \definecolor{codegreen}{rgb}{0,0.5,0.0}
\definecolor{codegray}{rgb}{0.5,0.5,0.5}
\definecolor{codepurple}{rgb}{0.58,0,0.82}
\definecolor{backcolour}{rgb}{0.95,0.95,0.92}

\newcommand*{\myfont}{\fontfamily{lmtt}\selectfont}

\newcommand{\commentKhaled}[1]{{\color{black}{#1}}}

\newcommand{\toolname}{{BatchLayout}}
\newcommand{\toolnameBH}{{BatchLayoutBH}}

\makeatletter
\def\BState{\State\hskip-\ALG@thistlm}
\makeatother

\makeatletter
\newcommand\fs@norules{\def\@fs@cfont{\bfseries}\let\@fs@capt\floatc@ruled
  \def\@fs@pre{}%
  \def\@fs@post{}%
  \def\@fs@mid{\kern3pt}%
  \let\@fs@iftopcapt\iftrue}
\makeatother
\floatstyle{norules}
\restylefloat{algorithm}

\ifpdf
  \pdfoutput=1\relax                   
  \usepackage{graphicx}                
  \DeclareGraphicsExtensions{.pdf,.png,.jpg,.jpeg} 
\else
  \usepackage{graphicx}                
  \DeclareGraphicsExtensions{.eps}     
\fi%

\graphicspath{{figures/}{pictures/}{images/}{./}} 

\usepackage{microtype}                 
\PassOptionsToPackage{warn}{textcomp}  
\usepackage{textcomp}                  
\usepackage{mathptmx}                  
\usepackage{times}                     
\usepackage{cite}                      
\usepackage{tabu}                      
\usepackage{booktabs}                  

\onlineid{0}

\vgtccategory{Research}

\newcommand\correspondingauthor{\thanks{Corresponding author.}}

\usepackage{xcolor}
\def\BibTeX{{\rm B\kern-.05em{\sc i\kern-.025em b}\kern-.08em
    T\kern-.1667em\lower.7ex\hbox{E}\kern-.125emX}}
\begin{document}
\setcounter{page}{1}
\title{BatchLayout: A Batch-Parallel Force-Directed \\Graph Layout Algorithm in Shared Memory\\
}

\author{Md. Khaledur Rahman\thanks{e-mail: morahma@iu.edu, Department of Computer Science}\\ 
        \scriptsize Indiana University Bloomington %
\and Majedul Haque Sujon\thanks{e-mail: msujon@iu.edu, Department of Intelligent Systems Engineering} \\ \scriptsize  Indiana University Bloomington
\and Ariful Azad \correspondingauthor\; 
\thanks{e-mail: azad@iu.edu, Department of Intelligent Systems Engineering}\\ 
     \scriptsize  Indiana University Bloomington}

\abstract{

Force-directed algorithms are widely used to generate aesthetically-pleasing  layouts of graphs or networks arisen in many scientific disciplines.
To visualize large-scale graphs, several parallel algorithms have been discussed in the literature.  
\commentKhaled{However, existing parallel algorithms do not utilize memory hierarchy efficiently and often offer limited parallelism.}
This paper addresses these limitations with \toolname{}, an algorithm that groups vertices into minibatches and processes them in parallel. 
\toolname{} also employs cache blocking techniques to utilize memory hierarchy efficiently.
More parallelism and improved memory accesses coupled with force approximating techniques, better initialization, and optimized learning rate make \toolname{} significantly faster than other state-of-the-art algorithms such as ForceAtlas2 and OpenOrd.
The visualization quality of layouts from \toolname{} is comparable or better than similar visualization tools.
All of our source code, links to datasets, results and log files are available at \url{https://github.com/khaled-rahman/BatchLayout}.}


\CCScatlist{
  \CCScatTwelve{Human-centered computing}{Visu\-al\-iza\-tion}{Visu\-al\-iza\-tion techniques}{Graph drawings};
  \CCScatTwelve{Human-centered computing}{Visu\-al\-iza\-tion}{Visu\-al\-iza\-tion
  systems and tools}{Visualization toolkits}{}
}
\maketitle

\section{Introduction}

Networks or graphs are common representations of scientific, social and business data.
In a graph, a set of vertices represents entities (e.g., persons, brain neurons, atoms) and a set of edges indicates relationships among entities (friendship, neuron synapses, chemical bonds).
A key aspect of data-driven graph analytics is to visually study large-scale networks, such as biological and social networks, with millions or even billions of vertices and edges. \commentKhaled{In network visualization, the first step is to generate a layout in a 2D or 3D coordinate system which can be fed into visualization tools such as Gephi~\cite{bastian2009gephi} and Cytoscape~\cite{shannon2003cytoscape}}. 
Therefore, the quality and computational complexity of network visualization are often dominated by the graph layout algorithms. 

Force-directed layout algorithms are among the most popularly used techniques to generate the layout of a graph. Following the philosophy of the spring energy model, these algorithms calculate attractive and repulsive forces among vertices in a graph and iteratively minimize an energy function. Classical force-directed algorithms, such as the Fruchterman and Reingold (FR) algorithm~\cite{fruchterman1991graph}, 
require $O(n^2)$ time per iteration where $n$ is the number of vertices in a graph. 
By approximating the repulsive force between non-adjacent nodes, we can get a faster $O(n\log n)$ algorithm. 
In this paper, we used the Barnes-Hut approximation~\cite{barnes1986hierarchical} based on the quad-tree data structure. 
\commentKhaled{Even though the layout quality (measured by stress,  neighborhood preservation, and other well established quality metrics as shown in Table~\ref{tab:measures_st_np}) from the Barnes-Hut approximation can be worse than an exact $O(n^2)$ algorithm, the former is often used to visualize large-scale networks. 
In this paper, we carefully analyzed the quality and runtime of both exact and approximate force-directed algorithms. 
}

\commentKhaled{
To visualize large graphs, several prior work discussed parallel algorithms for multicore processors~\cite{martin2011openord}, graphics processing unit (GPU)~\cite{jacomy2014forceatlas2, brinkmann2017exploiting}, and distributed clusters~\cite{arleo2017large,arleo2018distributed}. 
While we summarize state-of-the-art algorithms and software in the Related Work section, our primary focus lies in parallel algorithms for multicore servers. 
}
\commentKhaled{Some state-of-the-art force-directed algorithms such as ForceAtlas2~\cite{jacomy2014forceatlas2} and OpenOrd~\cite{martin2011openord} have parallel implementations for multicore processors. 
However, these algorithms do not utilize full advantage of present memory hierarchy efficiently and offer limited parallelism because they calculate forces for one vertex at a time. 
We aim to improve these two aspects of parallel algorithms}. We develop a parallel algorithm called \emph{BatchLayout} that offers more parallelism by grouping vertices into minibatches and processing all vertices in a minibatch in parallel.
This approach of increasing parallelism via minibatches is widely used in training deep neural networks in the form of minibatch Stochastic Gradient Descent (SGD) ~\cite{goyal2017accurate}. 
We adapt this approach to graph layout algorithms for better parallel performance.

Existing algorithms also access memory irregularly when processing sparse graphs with skewed degree distributions.
Irregular memory accesses do not efficiently utilize multi-level caches available in current processors, impeding the performance of algorithms.       
In BatchLayout, we regularized memory accesses by using linear algebra operations similar to matrix-vector multiplication and employing  ``cache blocking" techniques to utilize memory hierarchy efficiently.  
\commentKhaled{More parallelism and better memory accesses made BatchLayout faster than existing  parallel force-directed algorithms}. 
On average, the exact $O(n^2)$ version of BatchLayout is $15\times$ faster than ForceAtlas2.
The Barnes-Hut approximation in BatchLayout is about $4\times$ and $2\times$ faster than ForceAtlas2 (with BH approximation) and OpenOrd, respectively (the quality of an OpenOrd layout is usually worse than BatchLayout and ForceAtlas2).

BatchLayout attains high performance without sacrificing the quality of the layout. 
We used \commentKhaled{three} aesthetic metrics to quantitatively measure the quality of layouts and also visualized the networks in Python for human verification. 
According to all quality measures, BatchLayout generates similar or better-quality layouts than ForceAtlas2 and OpenOrd.

Overall, BatchLayout covers all important aspects of force-directed layout algorithms. 
We investigated four energy models, two initialization techniques, different learning rates and convergence criteria, various minibatch sizes, and \commentKhaled{three} aesthetic metrics to study the performance and quality of our algorithms.
All of these options are available in an open source software that can be run on any multicore processor.  
Therefore, similar to ForceAtlas2, BatchLayout provides plenty of options to users to generate a readable layout for an input graph. 

In this paper, we present new insights on parallel force-directed graph layout algorithms. We summarize key contributions below:
\begin{itemize}[leftmargin=*]
    \setlength\itemsep{0.02em}
    \item {\bf Improved parallelism:} We develop a class of exact and approximate parallel force-directed graph layout algorithms called \toolname{}. Our algorithms expose more parallel work to keep hundreds of processors busy.
    \item {\bf Better quality layouts:}  \toolname{} generates layouts with better or \commentKhaled{similar} qualities (according to various metrics) compared to other force-directed algorithms. 
    \item {\bf Comprehensive coverage:} \toolname{} provides multiple options for all aspects of force-directed algorithms. Especially, we cover new initialization techniques and four energy models.
    \item {\bf High-performance implementation:} We provide an implementation of \toolname{} for multicore processors. 
    \toolname{} improves data locality and decreases thread synchronizations for better performance. 
    \item {\bf Faster than ForceAtlas2 and OpenOrd:} On a diverse set of graphs, \toolname{} is significantly faster than ForceAtlas2 and OpenOrd.  
    
\end{itemize}


\section{Background}
\subsection{Notations}
We present an unweighted graph by ${G=(V,E)}$ on the set of vertices $V$ and set of edges $E$. 
The number of vertices and edges are denoted by $n$ and $m$, respectively.
In this paper, we only considered undirected graphs, but directed graphs can be used by ignoring the directionality of the edges. 
The degree of vertex $i$ is denoted by $deg(i)$.
We denote the coordinate of vertex $i$ with $c_i$, which represents $(x,y)$ Cartesian coordinate in a $2D$ plane. The distance between two vertices $i$ and $j$ is represented by $\parallel c_i - c_j \parallel$.

In our algorithm, we used the standard Compressed Sparse Row (CSR) data structure to store the adjacency matrix of a sparse graph. 
For unweighted graphs, the CSR format consists of two arrays: $rowptr$ and $colids$. $rowptr$, an array of length $n+1$, stores the starting indices of rows. $colids$ is an array of length $m$ that stores the column indices in each row. 
Here, the CSR data structure is used for space and computational efficiency, but any other data structure would also work with our algorithm.


\subsection{Force Calculations}
Force-directed graph drawing algorithms compute \emph{attractive} forces between every pair of adjacent vertices and \emph{repulsive} forces between nonadjacent vertices. 
In the spring-electrical model, a pair of connected vertices attract each other like a spring based on Hooke's law and a pair of nonadjacent vertices repulse each other like electrically charged particles based on Coulomb's law.
Following this spring-electrical model, we define the attractive force $f_a(i,j)$ and the repulsive force $f_r(i,j)$ between vertices $i$ and $j$ as follows:
\begin{equation*}
    \vspace{-0.2cm}
    f_a(i,j) = \frac{\parallel c_i - c_j \parallel^a}{K}, \text \  \ 
    f_r(i,j) = \frac{-RK^2}{\parallel c_i - c_j \parallel^r}.
\end{equation*}
Here, $K$ acts as an optimal spring length and $R$ is considered as the regulator of relative strength of forces. Different values of $a$ and $r$ give different energy models, known as the $(a,r)$-energy model \cite{jacomy2014forceatlas2,noack2009modularity}. The combined force $f(i, c)$ of a vertex $i$ with respect to the current coordinates $c$ of all vertices is computed by:
\vspace{-0.2cm}
\begin{equation*}
    f(i, c) = \sum_{\mathclap{(i,j) \in E}}f_a(i,j)\times \frac{c_j - c_i}{\parallel c_i-c_j \parallel} + \sum_{\mathclap{\substack{i\neq j \\ (i,j) \notin E} }} {f_r(i,j)} \times \frac{c_j - c_i}{\parallel c_i-c_j\parallel}.
    \vspace{-0.3cm}
\end{equation*}
Here, $c$ is an array of coordinates of all vertices i.e., $c = \{c_i|i\in V\}$.
We multiplied attractive and repulsive forces by a unit vector to get the direction of movement of vertex $i$. 
Hence, following the spring-electrical model, the \emph{energy} of the whole graph will be proportional to $ \sum_{i\in V} f^2(i, c) $. 
The goal of a force-directed algorithm is to minimize the energy so that the layout of the graph becomes stable, readable and visually appealing.

\begin{algorithm}[t]
\caption{A Sequential Force-Directed Algorithm}\label{euclid}
\hspace*{\algorithmicindent} \textbf{Input:} G(V, E) and an initial layout $c$
\begin{algorithmic}[1]
\State $Step = 1.0$ \Comment{initial step length}
\State $Loop = 0$
\While {$ Loop < MaxIteration$}
\State $Energy = 0$
\For{$i \leftarrow 0$ to $n -1$}
    \State $f = 0$ \Comment{force initialization}
        \For{$j \leftarrow 0$ to $n-1$}
            \If{$(i,j)\in E$}
                \State $f= f+ f_a(i,j)\times \frac{c_j - c_i}{\parallel c_i - c_j\parallel}$
            \Else 
                \State $f = f + f_r(i,j)\times \frac{c_j - c_i}{\parallel c_i - c_j\parallel}$
            \EndIf
        \EndFor
        \State $c_i = c_i + Step \times \frac{f}{\parallel f\parallel}$ 
        \State $Energy = Energy + \parallel f \parallel^2$
    \EndFor
\State $Step = Step \times 0.999$
\State $Loop = Loop + 1$
\EndWhile
\State \Return the final layout $c$
\end{algorithmic}
\label{algo:sequential}
\vspace{-0.7cm}
\end{algorithm}

\begin{figure*}[h]
    \centering
    \includegraphics[width=0.9\linewidth]{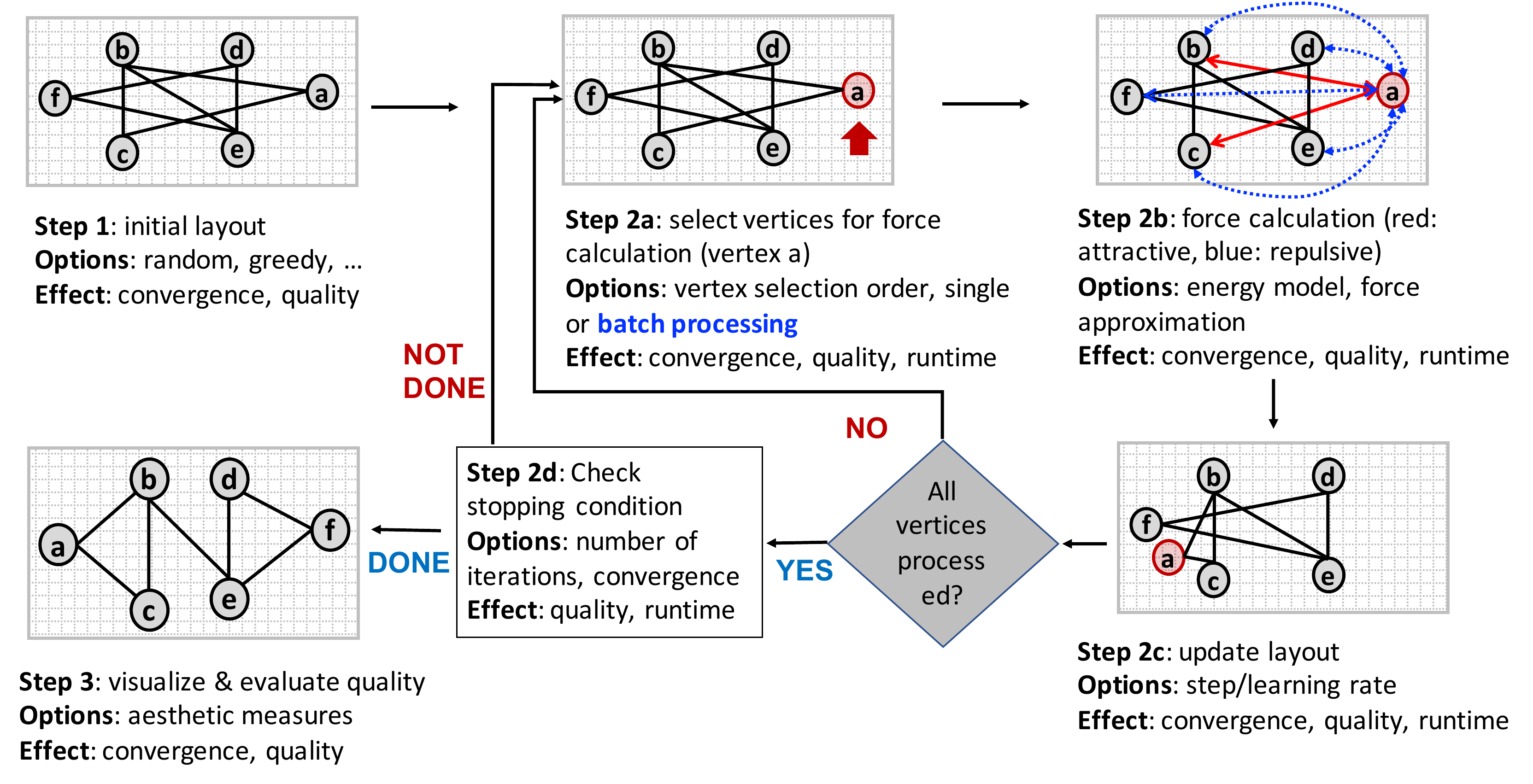}
    \vspace{-0.5cm}
    \caption{Steps involved in a typical force directed graph layout algorithm. An algorithm starts with an initial layout (usually a random layout) and iteratively updates the layout by calculating attractive and repulsive forces among vertices. When a stopping criterion is satisfied, the graph is visualized using the generated layout and the quality of the layout is evaluated. Each step has various algorithmic options that impact the convergence speed, the quality of the final layout, and the runtime of the algorithm. We highlighted ``batch processing" in Step 2a since it is a required option for efficient parallel algorithms discussed in this paper. We explored many options within our parallel algorithm.}
    \vspace{-0.4cm}
    \label{fig:batchlayoutsystem}
\end{figure*}

\subsection{A Standard Sequential Algorithm}
Algorithm~\ref{algo:sequential} describes a standard sequential force-directed graph layout algorithm. 
We closely followed the notation of an Adaptive Cooling Scheme by Y. Hu~\cite{hu2005efficient}. 
Algorithm~\ref{algo:sequential} starts with an initial layout and iteratively improves it by minimizing energy.
In each iteration, vertices are chosen in a predefined order (line 5), and attractive and repulsive forces are calculated for a selected vertex with respect to all other vertices in the graph.
Next, the relative position of the selected vertex is updated (line 12) and this updated position is used in processing the subsequent vertices. 
\emph{Step} used in line 12 is a hyper-parameter that influences the convergence of the algorithm.  
We will discuss possible values of this hyper-parameter in the experimental section.
A working example of this algorithm is shown in Fig. S1 of the supplementary file. 
Each iteration of Algorithm~\ref{algo:sequential} computes $O(m)$ attractive forces and $O(n^2)$ repulsive forces, giving us a sequential $O(n^2)$ algorithm.
Using the Barnes-Hut approximation~\cite{barnes1986hierarchical} of repulsive forces, Algorithm~\ref{algo:sequential} can easily be made into a sequential $O(n\log n)$ algorithm. 

\section{The \toolname{} Algorithm}

\subsection{Toward A Scalable Parallel Algorithm}
Despite its simplicity, it is hard to yield high performance from Algorithm~\ref{algo:sequential}.
A na\"ive parallelization of Algorithm~\ref{algo:sequential} can simply compute forces of a vertex in parallel by making line 7 of Algorithm~\ref{algo:sequential} a parallel loop. 
However, for vertex $i$, this na\"ive parallel algorithm offers $O(n+deg(i))$ parallel work for the exact algorithm and $O(\log n+deg(i))$ parallel work for the approximate algorithm.
To increase the parallel work, we followed the trend in training deep neural networks via SGD.
In traditional SGD, the gradient is calculated for each training example, which is then used to update model parameters.
Since SGD offers limited parallelism, most practical approaches use a batch of training examples (batch size varies from 256 to several thousands) to compute the gradient, an approach known as minibatch SGD.
We follow a similar approach by calculating forces for a batch of vertices and updating their coordinates in parallel. 
\toolname{} revolves around the batch processing of vertices and covers all other aspects of force-directed algorithms.

\subsection{Overview of \toolname}
Figure~\ref{fig:batchlayoutsystem} shows the skeleton of a typical force-directed graph layout algorithm. 
After starting with an initial layout, an algorithm iteratively minimizes energy by computing attractive and repulsive forces.
Upon convergence, the final layout is drawn using a visualization tool. 
Figure~\ref{fig:batchlayoutsystem} shows that different steps of the algorithm have various choices upon which the quality and runtime of the algorithm depend. 
A comprehensive software such as ForceAtlas2 provides multiple options in each step. 
Like ForceAtlas2, \toolname{} also provides multiple options in each of these steps.
In particular, \toolname{} uses two initialization strategies, process vertices in different orders, considers four energy models, calculates exact and approximate repulsive forces, explores different learning rates and convergence conditions.
We evaluated the impact of different options on the convergence and quality of the algorithm.
We developed parallel algorithms for all options and evaluated their parallel performance.

\subsection{Batch processing of vertices}
\begin{figure}[!htb]
    \centering
    \includegraphics[width=\linewidth]{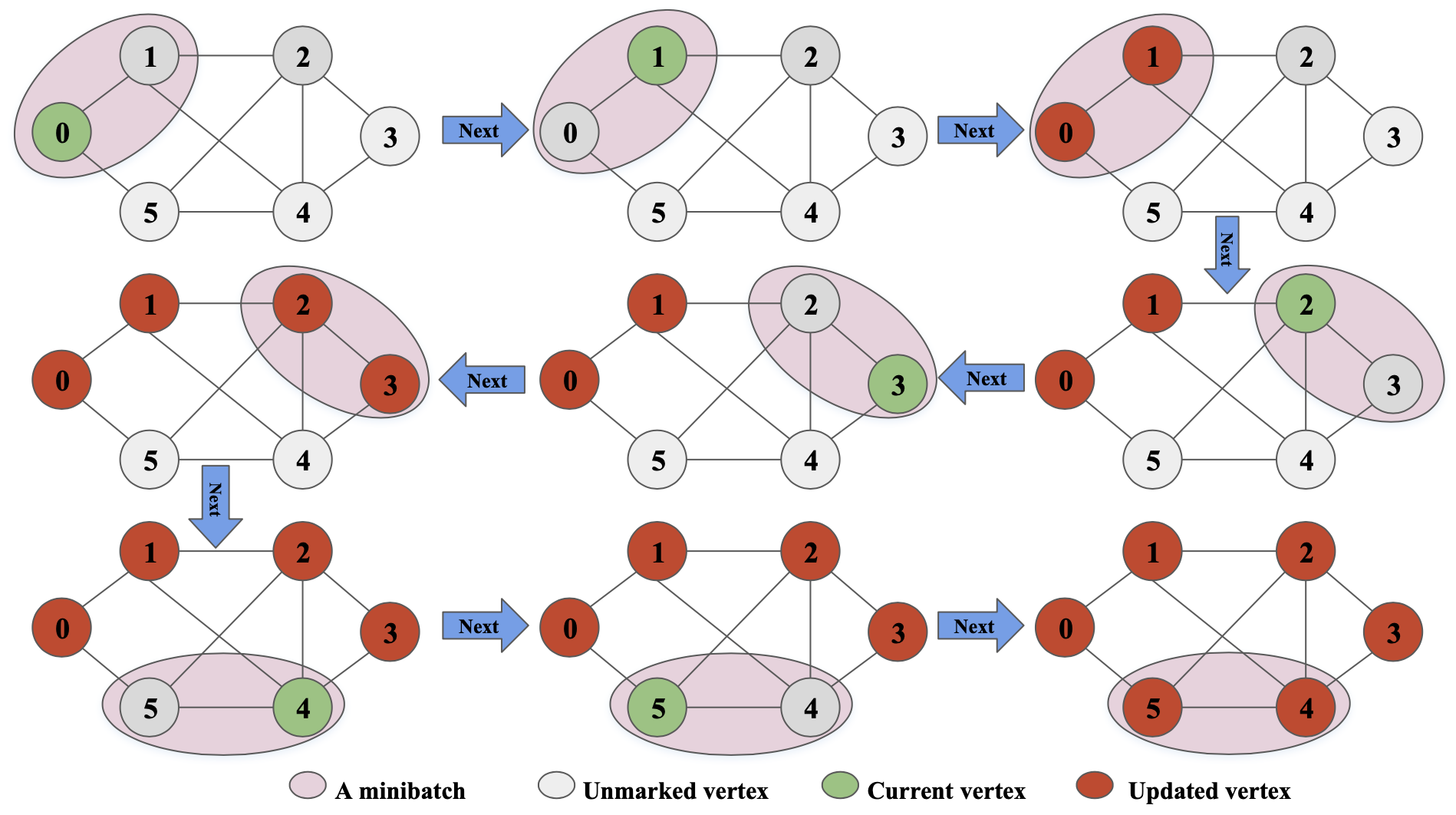}
    \caption{Force calculation with minibatch size 2. We start from vertex $0$ and consider two consecutive vertices in a minibatch shown in shaded regions. 
    This example finishes in three steps (each row in the figure), and each step processes a minibatch of two vertices.
    Green color denotes a vertex whose force is currently being computed, and red denotes a vertex whose coordinate is being updated. 
    In the first step (first row), vertex $0$ and vertex $1$ form a minibatch whose forces are first computed and then their coordinates are updated. 
    Similarly, in the second step (second row, from right to left), we calculate forces for vertex $2$ and $3$, and then update their coordinates. 
    Notice that in step 2, updated coordinates of the previous minibatch $\{0, 1\}$ are used.
    }
    \vspace{-0.5cm}
    \label{fig:BatchLayoutfig}
\end{figure}

Since force calculations consume almost all computing time of our algorithm, we discuss parallel force calculation techniques first. 
As mentioned before, calculating forces for a single vertex does not provide enough work to keep many processors busy.
Hence, the \toolname{} algorithm selects a subset of vertices called a \emph{minibatch} and calculates forces for each vertex in the minibatch independently.
After all vertices in a minibatch finish force calculations, we update their coordinates in parallel.
Therefore, unlike Algorithm~\ref{algo:sequential}, the minibatch approach delays updating coordinates until all vertices in a minibatch finish their force calculations.
Fig. \ref{fig:BatchLayoutfig} illustrates this approach with a minibatch size of 2.

Batched force calculation has a benefit and a drawback. 
As we process more vertices simultaneously, we can keep more processors busy by providing ample parallel work. 
However, batch processing may slow down the convergence because the updated coordinates of a vertex cannot be utilized immediately. 
In our experiment, we observed that the benefit of concurrent work is significantly more than the disadvantage caused by slower convergence. 
Hence, batch processing makes a parallel algorithm significantly faster. 

\begin{figure}
    \centering
    \includegraphics[width=0.7\linewidth]{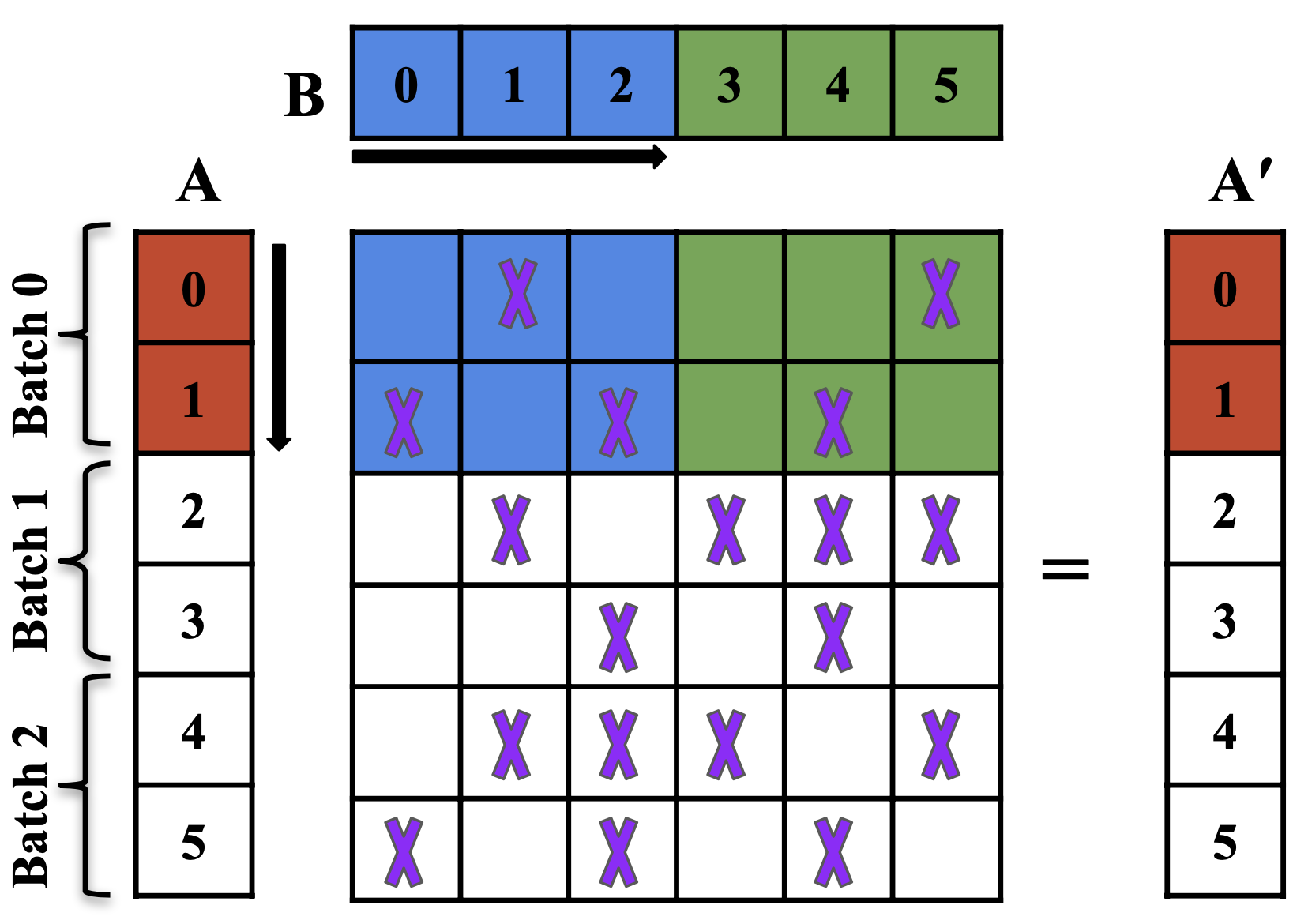}
    \vspace{-0.2cm}
    \caption{Computation of forces for the graph in Fig.~\ref{fig:BatchLayoutfig} using cache blocking technique. Here, the graph is stored as an adjacency matrix with `X' denoting edges between vertices. Vectors represents coordinates of vertices, and force calculations are viewed as a matrix-vector multiplication.
    }
    \vspace{-0.3cm}
    \label{fig:cacheblocking}
\end{figure}

\subsection{Improved Memory Accesses Via Cache Blocking}
While batch processing of vertices provides ample parallel work, processors can still remain idle if they have to wait for data to be fetched from memory.  
Memory stalling can happen if an algorithm does not utilize spatial and temporal locality in accessing data.
We addressed this issue by using cache blocking, a technique frequently used in linear algebra operations to utilize the cache hierarchy available in modern processors. 




We noticed that force calculations can be viewed as a matrix-vector multiplication. 
Fig. \ref{fig:cacheblocking} shows a simple example where the adjacency matrix of the graph used in Fig.~\ref{fig:BatchLayoutfig} is considered.
The vector of current coordinates are shown twice in dense vectors A and B.
$A'$ holds the updated coordinates.
In our minibatch model,  we compute the forces for all vertices in a minibatch with respect to all other vertices in the graph.
Suppose, in batch $0$, we compute forces for vertices $0$ and $1$ with respect to vertices $0 \ldots 2$ (blue). When forces are calculated for vertex $0$, coordinates of vertices $0 \ldots 2$ are brought in from the main memory and stored in a CPU cache.
The cumulative sum of force calculations are stored in a temporary vector, $A'$. 
When we compute forces for the next vertex $1$, all necessary coordinates are already available in cache (i.e., using data locality), which reduces the memory traffic. Batch $0$ gets a similar performance advantage when forces are computed with respect to vertices $3\ldots 5$ (green). 
\toolname{} with the cache-blocking scheme is presented in Algorithm \ref{algo:cbBatchLayout}. 
Here, {\myfont G.rowptr} and {\myfont G.colids} are two arrays which hold the row pointers and column IDs, respectively, of the graph in a CSR format. 
$BS$ is the size of minibatch and $Step$ length is initialized to $1.0$. 
In line $6$ of Algorithm \ref{algo:cbBatchLayout}, we iterate over a subset of vertices (minibatch) and in line $10$, we iterate over all vertices to calculate $f_a$ and $f_r$. 
In line $14$, when two vertices are connected by an edge we calculate $f_a$; otherwise, we calculate $f_r$ (line $18$). After calculating forces for a vertex, we store it in a temporary location (see line $19$). 
This temporary storage helps us to calculate forces in parallel. 
Next, we iterate over all vertices in a minibatch to update relative position (see line $22$).  
We provide a flexible rectangular cache blocking area which is bounded by variables $p$ and $q$. 
This increases the data locality in the L1 cache of each core, improving the performance of our algorithm. 
Loop counters $i$ and $j$ are increased by $p$ and $q$, respectively. 
Variable $k_x$ holds the starting index of neighbors for $p$ vertices (see line $8$). After the loop completes, in line $24$, we update step length. 
To keep Algorithm~\ref{algo:cbBatchLayout} simple, we assume that the number of vertices in a graph is multiple of $(b+1)*BS$.
In our implementation, we incorporated other cases.
The overall running time of this procedure is same as Algorithm \ref{euclid} i.e., $O(n^2)$. However, the main advantage of this technique is in parallel computation of forces. 
Note that Algorithm \ref{euclid} is a special case of Algorithm \ref{algo:cbBatchLayout} when the size of the minibatch is $1$.

\begin{algorithm}
\caption{Cache Blocking \toolname}
\hspace*{\algorithmicindent} \textbf{Input:} G(V, E)
\begin{algorithmic}[1]
\State $Step = 1.0$ \Comment{initial step length}
\State $Loop = 0$
\While {$ Loop < MaxIteration$}
\State $Energy = 0$
\For{$b \leftarrow$ 0 to $\frac{n}{BS}-1$}
    \ParFor{$i \leftarrow b * BS$ to $(b + 1) * BS-1$ \textbf{by} $p$}
        \For{$x \leftarrow 0$ to $p-1$}
            \State $k_{x} = G.rowptr[i+x]$
			\State $ct_{i+x} = 0$
        \EndFor
        \For{$j \leftarrow 0$ to $n-1$ \textbf{by} $q$}
            \For{$x \leftarrow 0$ to $p-1$}
            \State $f = 0$
                \For{$y \leftarrow 0$ to $q-1$}
                \If{$j + y == G.colids[k_x]$}
                    \State $f\;+= f_a(i,j)\times \frac{c_j - c_i}{\parallel c_i - c_j\parallel}$
                    \State $k_x\;+=1$
                \Else 
                    \State $f\; += f_r(i,j)\times \frac{c_j - c_i}{\parallel c_i - c_j\parallel}$
                \EndIf
                \EndFor
                \State $ct_{i+x} += f$ 
            \EndFor
        \EndFor
    \EndParFor
    \For{$i \leftarrow b * BS $ to $(b + 1) * BS - 1$}
        \State $c_i\; += Step \times \frac{ct_i}{\parallel ct_i\parallel}$
        \State $Energy\; += \parallel ct_i\parallel^2$
    \EndFor
\EndFor
\State $Step = Step \times 0.999$
\State $Loop += 1$
\EndWhile
\State \Return the final layout $c$
\end{algorithmic}
\label{algo:cbBatchLayout}
\end{algorithm}

\subsection{Repulsive Force Approximation}
\begin{figure}[!t]
    \centering
    \includegraphics[width=\linewidth]{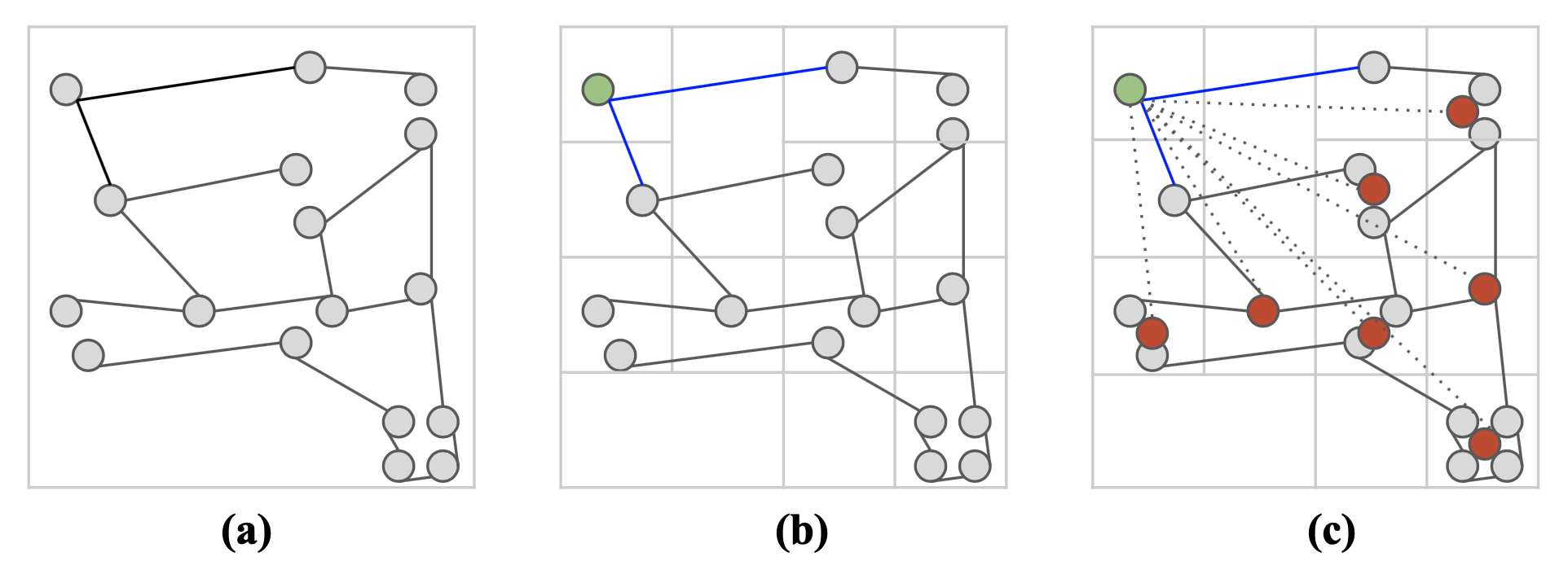}
    \vspace{-0.7cm}
    \caption{(a) An example graph. (b) Calculating forces of the green vertex using a quad-tree. (c) Blue solid lines show attractive forces, and dashed gray lines denote approximate repulsive forces with respect to centroids (red).}
    \vspace{-0.3cm}
    \label{fig:quadtree}
\end{figure}

Exact repulsive force computations need $O(n^2)$ time per iteration, which is too expensive for large graphs. 
Therefore, we have developed a parallel Barnes-Hut algorithm~\cite{barnes1986hierarchical} for approximating repulsive forces.
Barnes-Hut takes $O(n\log n)$ time per iteration. 
For 2D visualization, a quad-tree data structure is often used to approximate repulsive forces.
An example is shown in Fig.~\ref{fig:quadtree}. 
Given the coordinates of vertices in Fig.\ref{fig:quadtree}(a), a high-level block is divided into four equal blocks as shown in Fig. \ref{fig:quadtree}(b). 
In this example, we split up to $2$ levels, creating $4^2 = 16$ small blocks in the 2D plane. 
Empty blocks are merged with their adjacent nonempty blocks. 
We show how the algorithm approximates the repulsive force $f_r$ for the green vertex.
First, a centroid (shown in red) is calculated for each block by averaging the coordinates of all vertices in the block. 
Next, an approximate repulsive force is computed with respect to the centroids, as shown by the dashed gray lines. 

\begin{figure}[!t]
    \centering
    \includegraphics[width=\linewidth]{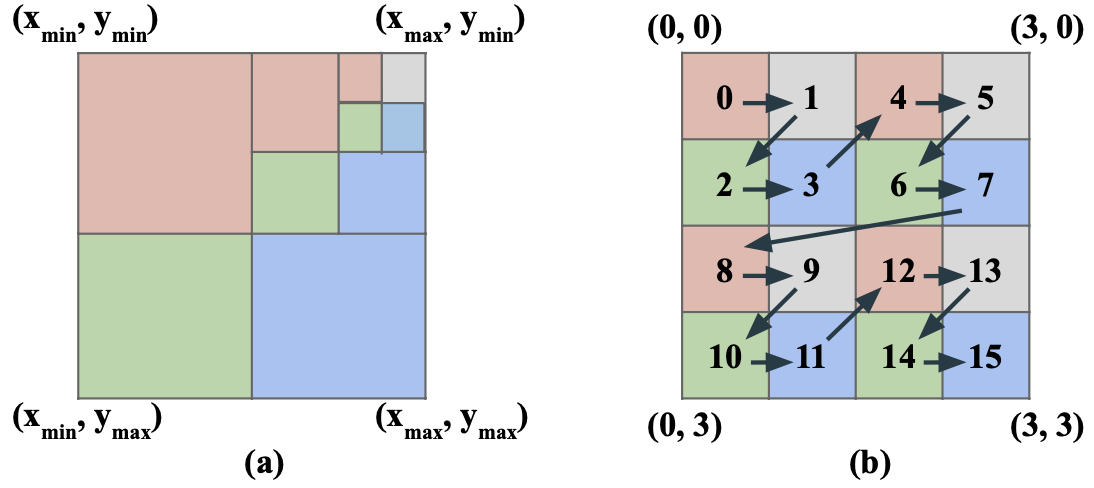}
    \vspace{-0.6cm}
    \caption{(a) A quad-tree structure. (b) Morton order of traversal for 16 vertices in 16 different blocks.}
    \vspace{-0.6cm}
    \label{fig:hashedquadtree}
\end{figure}

Among several choices of quad-tree implementation~\cite{zhang2014design}, we adopted an approach based on Warren and Salmon \cite{warren1993parallel}. 
Fig. \ref{fig:hashedquadtree}(a) shows that we define a drawing plane by bounding coordinates $(x_{max}, y_{max})$ and $(x_{min}, y_{min})$. 
This bounding plane is the root of the quad-tree.
Next, we split the bounding plane (root) into four equal blocks, each of which represents a child node (shown by a unique color) of the root. 
We sort the positions of all vertices based on Morton code (also called Z-order curve)~\cite{morton1966computer} and obtain an ordered list of vertices.
This ordering basically represents the Depth First Search (DFS) traversal of a quad-tree. 
Fig. \ref{fig:hashedquadtree}(b) shows a possible Morton ordering of a graph with 16 vertices. 
From the Morton ordering, we recursively build the tree, where leaf nodes represent vertices in the graph, and an internal node stores its centroid which is computed from its children.

Basically, nearer non-adjacent vertices contribute more than far vertices to $f_r$. For this reason, a threshold ($\theta$) is set to determine relative distance between two vertices while calculating $f_r$. 
A Multipole Acceptance Criteria (MAC), $\theta > \frac{D}{\parallel c_i - c_j \parallel}$ is evaluated while traversing the tree for $f_r$ calculations. If MAC is {\myfont\textbf{true}} then two vertices are located at far enough distance such that the current centroid can approximate that repulsive force. In this way, $f_r$ calculation can be approximated in $O(\log n)$ time which has been discussed elaborately \commentKhaled{in \cite{hu2005efficient,hachul2004drawing}}, thus, the overall running time becomes $O(n\log n)$ per iteration. We use the term \toolnameBH{} to describe this version and provide full pseudo-code in supplementary file, Algorithm S3.

\subsection{Initialization}
The initial layout plays an important role in the convergence and visualization quality of the final layout.
\commentKhaled{Many existing algorithms start with random layouts where vertex positions are assigned randomly~\cite{hu2005efficient}.}
In addition to random initial layouts, we developed a novel greedy initialization technique as shown in Algorithm \ref{algo:greedy}. 
We maintain a \emph{stack} $S$ to keep track of visited vertices.
The algorithm starts with a randomly selected vertex $V_0$, and $(0,0)$ is used as its coordinate. 
In every iteration, the algorithm extracts a vertex $U$ from the stack and places $U$'s neighbors on a unit circle centered at $U$ (line $10-14$ in Algorithm \ref{algo:greedy}).
Additionally, the distance between every pair of $U$'s neighbors is also equal.
In line $12$, $PI$ is a constant whose value is $3.1416$. 
In section \ref{subsec:uel}, we will discuss that uniform edge length is a good aesthetic metric for assessing the quality of a layout. 
Hence, we consider this unit-radius approach  to be a better initializer.
Our experimental results does show that the greedy initialization indeed converge faster than the random initialization. 
The greedy approach is similar to a DFS with $O(n+m)$ complexity.


\begin{algorithm}[!t]
\caption{Greedy Initialization}
\hspace*{\algorithmicindent} \textbf{Input:} G(V, E)
\begin{algorithmic}[1]
\State $S = \emptyset $ \Comment{initial empty stack}
\State $visited = [False, \ldots, False]$
\State $V_0 = (0, 0)$ \Comment{$V_{0x} = 0, V_{0y} = 0$}
\State $S.push(V_0)$
\State $visited_0 = True$
\While {$S$ is not empty}
\State $U = S.pop()$
\State $d = \frac{360}{deg(U)}$
\State $D = 0$
\For{$u$ neighbors of $U$}
    \If{$visited_u$ is $False$}
        \State $V_u = (U_x + cos(\frac{PI*D}{180.0}), U_y+sin(\frac{PI*D}{180.0}))$
        \State $visited_u = True$
        \State $S.push(V_u)$
        \State $D = D + d$
    \EndIf
\EndFor
\EndWhile
\State \Return $V$
\end{algorithmic}
\label{algo:greedy}
\vspace{-0.4cm}
\end{algorithm}

\subsection{Performance Metrics}
\label{sec:aesthetic_measure}
There are several aesthetic metrics in the literature to assess the quality of a layout generation algorithm \cite{purchase2002metrics,kwon2017would,de2019multi}. Though there does not exist a single metric that can uniquely determine the effectiveness of a layout, it gives a quantitative value that tells us how layout algorithms perform. For the sake of completeness, we have used the following aesthetic metrics along with running time analysis.


{\bf Stress (ST): }
Stress computes the difference between geometric distance and graph theoretic distance for any pair of vertices in a graph~\cite{brandes2008experimental,de2019multi}. Since different algorithms may require different drawing space based on which stress may change, the drawing space is scaled before computing the stress for all algorithms. Thus, the reported value is the minimum value achievable after scaling the layout. A lower ST value indicates a better layout, and it is computed using $\sum_{i,j\in V} w_{ij}(\parallel c_i - c_j\parallel - d_{ij})^2$ where, $c_i$ and $c_j$ are coordinates of vertices $i$ and $j$, respectively. The value of $d_{ij}$ represents a graph theoretic distance between vertices $i$ and $j$, and $w_{ij} = \frac{1}{d_{ij}^2}$.



{\bf Edge Uniformity (EU):}
\label{subsec:uel}
Sometimes the uniformity of edge lengths represents better readability of a layout \cite{huang2007effects}. We define EU in a similar way as defined in \cite{hachul2007large,de2019multi}. It finds the normalized standard deviation of edge length which can be computed using $\sqrt{\frac{\sum_{e\in E}(l_e - l_{\mu})^2}{|E|.l_{\mu}^2}}$ where, $l_{\mu}$ is the average length of all edges and $|E|$ represents the number of edges in the graph. Lower value of EU represents better quality of the layout.

{\bf Neighborhood Preservation (NP): }
This measure computes the number of neighbors that are close to a vertex in a graph as well as in the driven layout. It has been used to compare graph layouts in many tools like tsNET \cite{kruiger2017graph} and Maxent \cite{gansner2012maxent}. This is a normalized similarity measure where $0$ means that the adjacency of vertices are not preserved in the layout whereas $1$ means that all vertices preserve their respected neighbors.

\section{Results}

\subsection{Experimental Setup}
We implemented \toolname{} in C++ with multithreading support from OpenMP.
We ran all experiments in a server consisting of Intel Xeon Platinum 8160 processors (2.10GHz) arranged in two NUMA sockets. 
The system has 256GB memory, 48 cores (24 cores/socket), and 32MB L3 cache/socket. 
For comparison, we used ForceAtlas2 implemented in Gephi and OpenOrd from its  GitHub\footnote{https://github.com/SciTechStrategies/OpenOrd} repository.
We also experimented with Gephi's OpenOrd implementation (denoted by OpenOrdG), but found that the GitHub implementation produced better visualizations. 
We use Gephi's toolkit v0.9.2 \cite{bastian2009gephi} to run multi-threaded ForceAtlas2 and OpenOrdG. 
The Barnes-Hut variant of ForceAtlas2 is termed as ForceAtlas2BH. 
For \toolnameBH{}, we use the same threshold value ($1.2$) for MAC as in ForceAtlas2BH. 
Unless otherwise specified, we use the default settings for all of these tools.
For OpenOrd, we set the edge-cutting parameter to $0$, and we set the default time distribution in different stages of simulated annealing. 
While the \toolname{} software includes four variants of $(a,r)$-energy model considered in the ForceAtlas2 paper, we only show results from \toolname{} using $(2,-1)$-energy model (equivalent to the Fruchterman and Reigngold algorithm).
We selected this model because it usually generates better visualizations~\cite{jacomy2014forceatlas2}.

\begin{table}[]
\caption{Graphs used in our experiments. $|V|$ and $|E|$ represent number of vertices and edges, respectively. d means average degree.}
\vspace{-5pt}
\centering
\begin{tabular}{|p{1.35cm}|p{1.15cm}|p{1.34cm}|c|p{1.7cm}|}
\hline
\textbf{Graph} & \textbf{$|V|$} & \textbf{$|E|$} & \textbf{d} & \textbf{Type}\\ \hline
Powergrid             &      4,941	               &   13,188                &        	2.66              & Small World \\ \hline
add32            &     	4,960                &   	19,848                  & 	3.00       &   Scale Free  \\ \hline
ba\_network	           &      6,000	                &    5,999            &    	1.99                     & Scale Free \\ \hline
3elt\_dual	          & 9,000          & 	26,556	         &     2.95                  & Mesh \\ \hline
PGP           &         	10,680              &  	48,632	                &    4.55                 & Scale Free \\ \hline
pkustk02           &         	10,800             &      	399,600	             &        76              & Feiyue twin tower\\ \hline
fe\_4elt2	           &      11,143               &   	65,636	                & 5.89        &   Mesh  \\ \hline
bodyy6	           &        19,366              &    	134,208	                &        5.93             & NASA Mat.\\ \hline
pkustk01	            &       22,044               &         	979,380         &      	45.42               & Beijing bot. exhib. h. \\ \hline

OPF\_6000   &	29,902	&   274,697	    &   8.66    &  Ins. of Pow. Sys. G. U. \\ \hline
finance256	&   37,376	    &   298,496	    &   6.98    &   Lin. Prog. \\ \hline
finan512    &	74,752	    &   261,120	    &   6.98    &   Eco. Prob. \\ \hline
lxOSM	&   114,599 &	239,332 & 2.08 & Op. St. Map \\ \hline
comYoutube	&   1,134,890 &	5,975,248   & 5.26  & Social Net. \\ \hline
Flan\_1565	&   1,564,794	&   114,165,372   & 71.95  & Struc. Prob. \\ \hline
\end{tabular}
\label{tab:datasets}
\vspace{-0.5cm}
\end{table}

\subsection{Datasets}
\label{lab:datasets}
Table \ref{tab:datasets} shows a diverse collection of test graphs including small world networks, scale free networks, mesh, structural problem, optimization problem, linear programming, economic problem, social networks, and road networks. 
These graphs were collected from the SuiteSparse matrix collection (https://sparse.tamu.edu).
Specifically, we use Luxembourg OSM (lxOSM) and Flan\_1565 datasets to test the ability of algorithms to visualize large-scale graphs. 
For scalability experiments, we also generated benchmark networks by Lancichinetti et al.~\cite{lancichinetti2008benchmark}. 
For this tool, we set mixing parameter, avg. degree, community minsize, community maxsize to 0.034, 80, 128 and 300, respectively, and vary maximum degree from 100 to 300. 
We generated a total of 5 random graphs, each of which has $2^v$ vertices, where $v \in \{16, 17, 18, 19, 20\}$. 

\subsection{Algorithmic Options and Their Impacts}
\subsubsection{Initialization}
At first, we compare the differences between greedy initialization using Algorithm \ref{algo:greedy} and random initialization using the C++ function {\myfont rand}.
Both initialization techniques generate random coordinates within a given range {\myfont [-MAXMIN, MAXMIN]}.
We feed initial coordinates to \toolname{}, and Fig. \ref{fig:selfcheckout}(a) shows the layout energy for  the \emph{3elt\_dual} graph after various iterations. 
We observe that in the $i^{th}$ iteration, the greedy-initialized layout has lower energy than  the randomly-initialized layout.
This result is consistent across other graphs as well.
Fig. \ref{fig:selfcheckout}(a) reveals that greedy initialization accelerates the convergence of \toolname{}.
Hence,  we use the greedy initialization in subsequent experiments in the paper (both initializations are available in our software).

\subsubsection{Minibatch size}
\begin{figure*}[ht]
\includegraphics[width=0.33\linewidth]{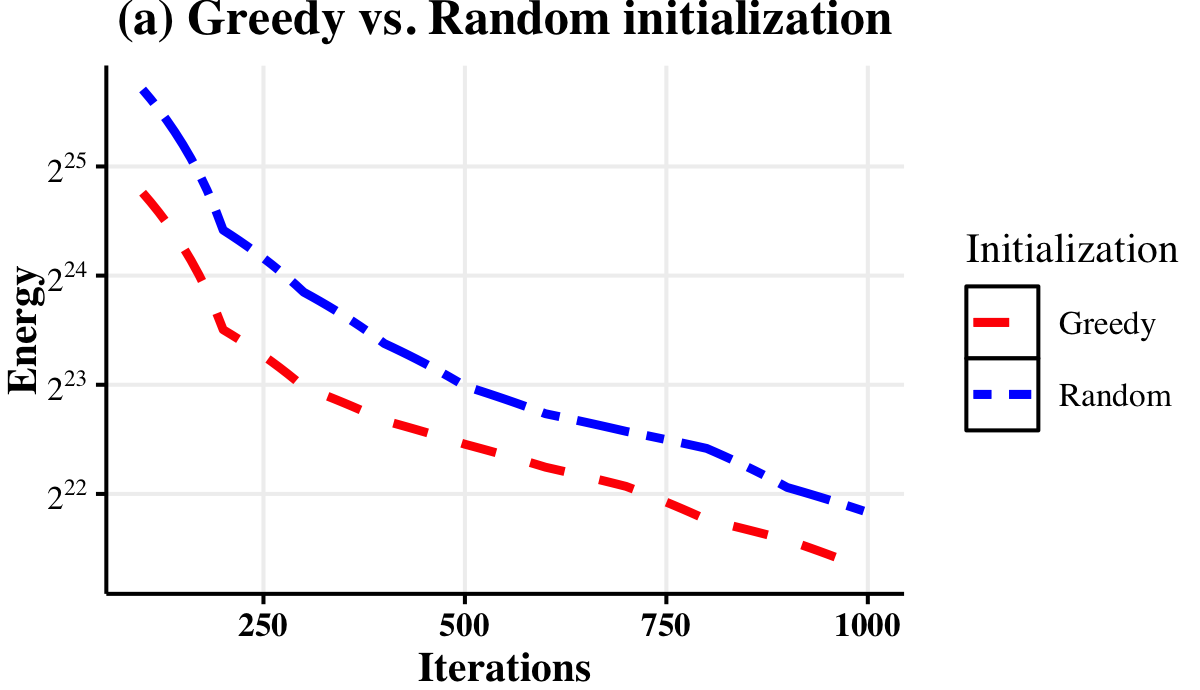}
\includegraphics[width=0.33\linewidth]{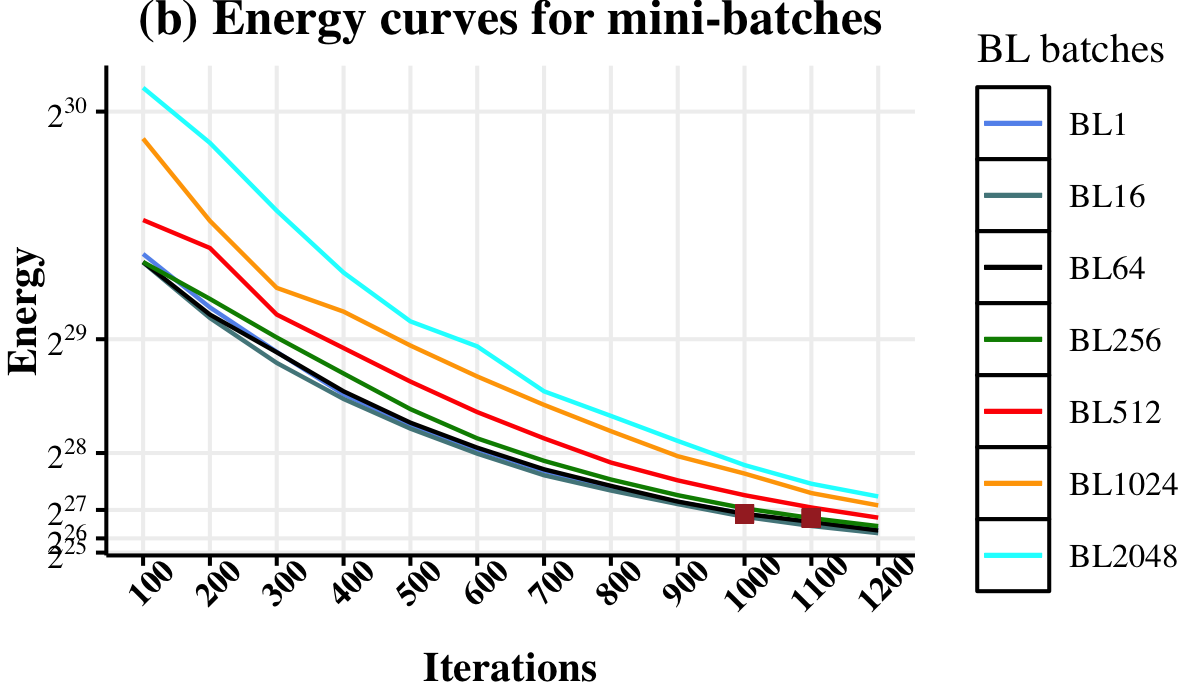}
\includegraphics[width=0.33\linewidth]{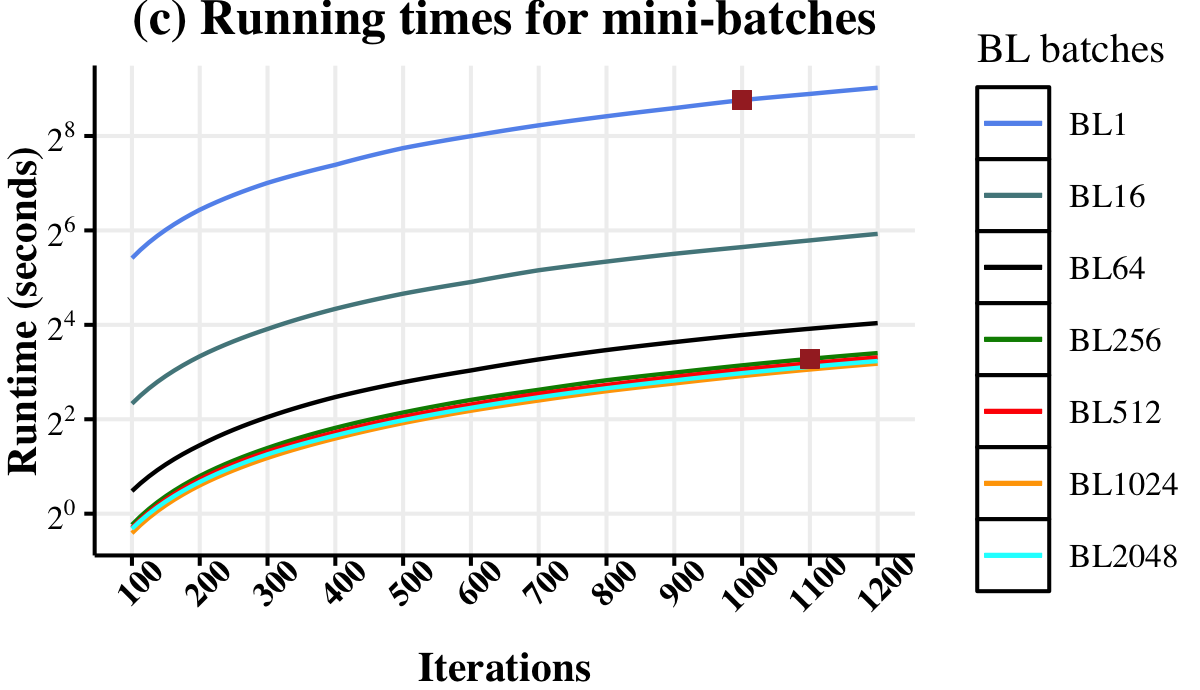}
\vspace{-0.4cm}
\caption{(a) Energy curves for greedy and random initialization (\emph{3elt\_dual} graph). (b) Energy curves ($\log$ scale) for different batches (\emph{pkustk02} graph). (c) Running time ($\log$ scale) for different batches of \toolname{} (\emph{pkustk02} graph).}
\label{fig:selfcheckout}
\end{figure*}

We now select a minibatch size that provides enough parallelism without affecting the convergence rate significantly.
Fig. \ref{fig:selfcheckout}(b) shows layout energies for different minibatches for the \emph{pkustk02} graph.
Here, BL256 denotes \toolname{} with minibatch size of 256.
For this graph, we observe that small minibatches converge faster (in terms of iterations) than large minibatches. 
For example, BL1 and BL256 achieve the same energy at 1000 and 1100 iterations, respectively (marked by brown squares in Fig. \ref{fig:selfcheckout}(b)). 
Consequently, BL256 needs to run 100 more iterations to reach the same energy achieved by BL1.
However, the cost of extra iterations is offset by faster parallel runtime of BL256.
Fig.~\ref{fig:selfcheckout}(c) shows the runtime for different minibatches when we run \toolname{} with 48 threads for the same graph \emph{pkustk02}.
To achieve the energy level marked by brown squares, BL1 takes around 433.6 seconds, whereas BL256 takes only 9.7 seconds.
That is, BL256 is $\sim 40\times$ faster than BL1 despite the former taking 100 more iterations than the latter to reach the same layout energy.
For this reason, minibatches play a central role in attaining high performance by our parallel algorithm.
However, using larger minibatches beyond 256 does not improve the performance further.
For example, BL256 and BL2048 behave almost similarly as can be seen in Figs. \ref{fig:selfcheckout}(b) and \ref{fig:selfcheckout}(c).
We also observe similar minibatch profiles for other graphs. Hence, we set 256 as the default minibatch size for the rest of our experiments (a user can change the minibatch size in our software).


\subsubsection{Batch Randomization}
We conduct experiments for the randomization of batches which is a common practice in machine learning. We reported standard deviation as well as energy curves for different runs using \emph{3elt\_dual} graph (Figure S3 in supplementary file). From our experiments, we observed that randomized batch selection achieves better energy curve than fixed sequential batch selection but results are not deterministic i.e., for each run, output is different (similar but not same) even though all hyper-parameters are same. We can see a greater difference for a smaller number of iteration. Randomization also introduces overhead in total running time for reshuffling and it can not take advantage of optimal cache usage. Moreover, randomized batch selection does not make any significant improvement in layout. So, we keep sequential batch selection as our default option in \toolname{}.

\subsubsection{Number of threads} 
Using more threads not necessarily translates to faster runtime, especially for small graphs.
This is because of the overhead of thread \emph{fork-join} model used in OpenMP.
For larger graphs with more than 100,000 vertices, we recommend employing all available threads.
For smaller graphs, users can consider using fewer threads. 
Hence, the number of threads is provided as an option in \toolname.


\begin{figure*}[ht]
\includegraphics[width=0.33\linewidth]{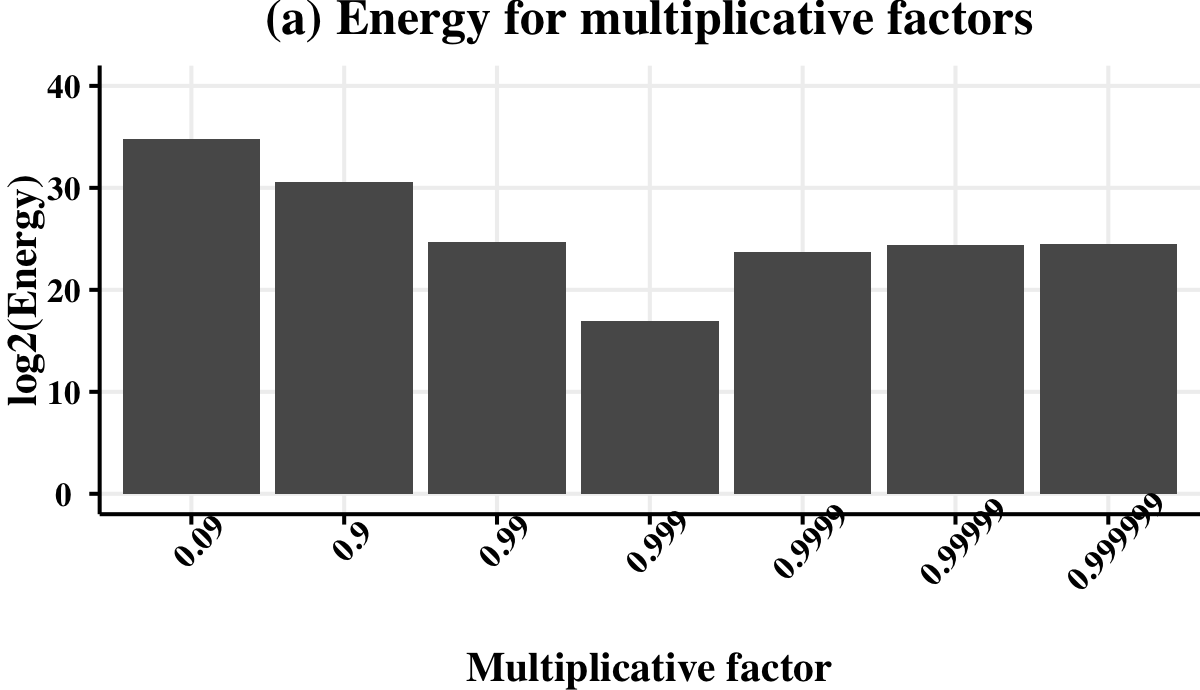}
\includegraphics[width=0.33\linewidth]{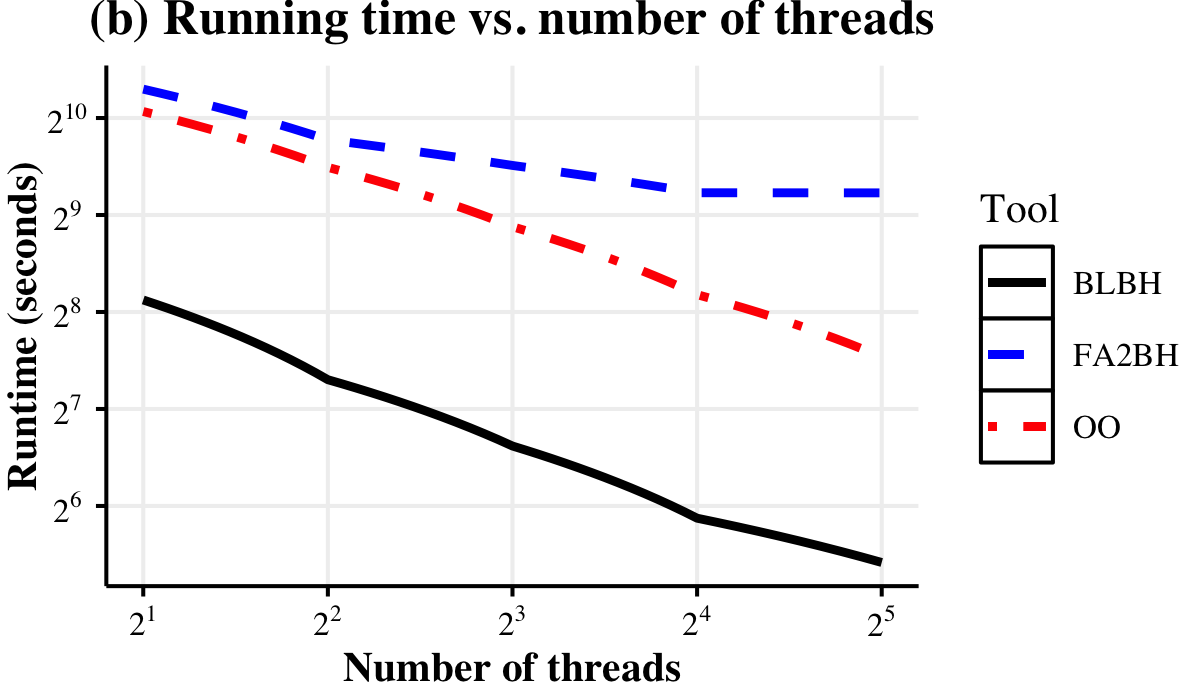}
\includegraphics[width=0.33\linewidth]{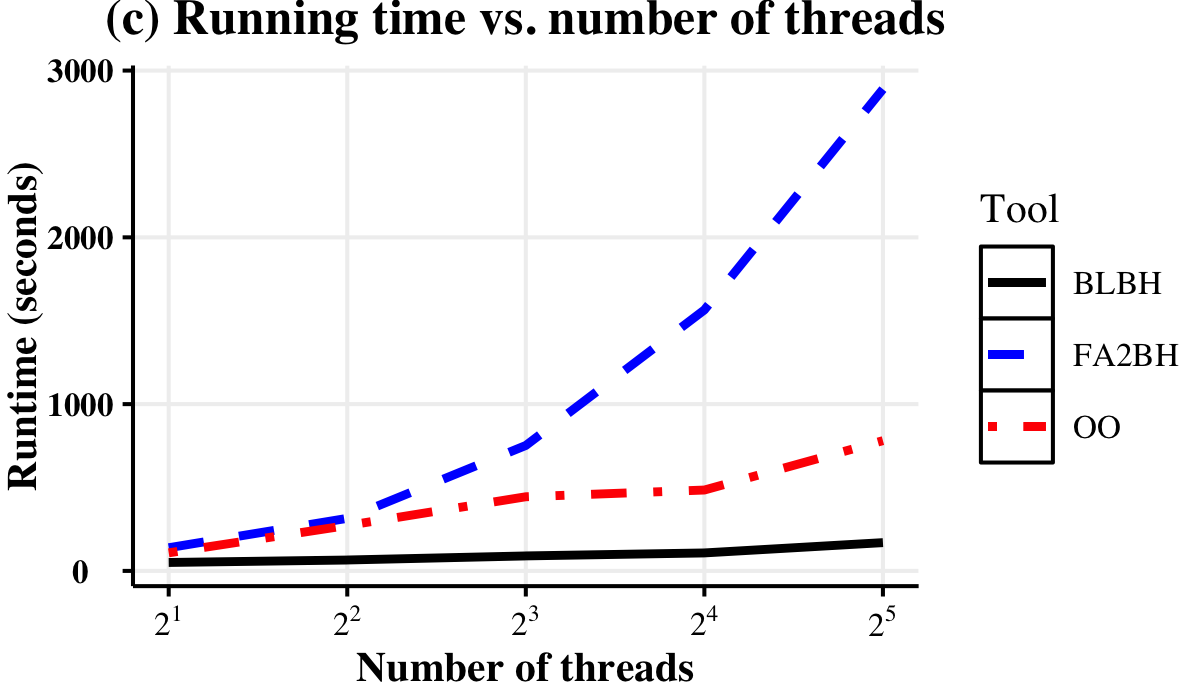}
\vspace{-0.4cm}
\caption{(a) Energy values ($\log$ scale) for different multiplicative factor of $Step$ using \emph{3elt\_dual} graph. (b) \commentKhaled{Strong scaling: Running time ($\log$ scale) for different number of threads using a synthetic graph of $2^{18}$ vertices generated by \cite{lancichinetti2008benchmark}. (c) Weak scaling: Running time (seconds) for different number of threads and graph sizes.}}
\vspace{-0.5cm}
\label{fig:selfcheckout2}
\end{figure*}

\subsubsection{Step length or learning rate}
Line 14 of Algorithm~\ref{euclid} uses a multiplicative factor for $Step$, which dictates both convergence and the quality of the final layout.
Intuitively, we want to adapt the step length so that the algorithm takes bigger steps in the beginning but increasingly smaller steps as we move closer to the minimum energy.
Fig.~\ref{fig:selfcheckout2}(a) shows layout energies at convergence with various multiplicative factors for the \emph{3elt\_dual} graph.
We observe that the factor of 0.999 (that is $0.1\%$ reduction of step length in every iteration) provides the best balance as the converged energy is minimum for this factor. 
Hence, we use this multiplicative factor in all experiments with \toolname{}.
We note that adaptive steps are also used in other force-directed layouts such as in the algorithm developed by Yifan Hu~\cite{hu2005efficient}.
In this paper, we empirically identify an effective adaption policy so that \toolname{} converges faster.


\subsection{Running Time and Scalability}
\label{lab:runtime}
We run all algorithms for 500 iterations using 47 threads (one thread per core) and report their runtime in Table~\ref{tab:runningtime}.
Since OpenOrdG leaves out a core for kernel-specific tasks, we also leave a core unutilized for fairness in comparison.  
In Table \ref{tab:runningtime}, we include time for all steps except file input/output operations.
If we consider exact force calculations, \toolname{} is on average $15\times$ faster than ForceAtlas2 (min:$7.6\times$, max:$21.8\times$).
For approximate force calculation, \toolnameBH{} is on average $2\times$ faster than OpenOrd (min:$1.4\times$, max:$9.5\times$).
Both \toolnameBH{} and OpenOrd are faster than ForceAtlas2BH.
Notice that OpenOrdG (that is Gephi's OpenOrd) also runs fast, but its layout quality is very low as seen in the next section.
Overall, \toolnameBH{} is almost always the fastest algorithm as denoted by bold numbers in Table \ref{tab:runningtime}.
We also tested tsNET~\cite{kruiger2017graph} to conduct some experiments (with perplexity and learning rate set to 800 and 6000, respectively). 
We observe that tsNET can generate good layouts for small graphs, but failed to generate any layout for medium or bigger graphs in our experiments. 
For small graphs like \emph{3elt\_dual}, tsNET took around 1 hour whereas other multi-threaded tools generated layout within few seconds. 
Hence, we did not include tsNET runtimes in Table \ref{tab:runningtime}. 

{\bf Thread and vertex scalability.} We now discuss the thread and vertex scalability of the three fastest algorithms: \toolnameBH{} (FLBH), ForceAtlas2BH (FA2BH), and OpenOrd (OO).
As discussed in section \ref{lab:datasets}, we use a benchmark tool~\cite{lancichinetti2008benchmark} to generate graphs for the scalability experiment.
Fig.~\ref{fig:selfcheckout2}(b) shows the thread scalability for a graph with $2^{18}$ vertices.
We observe that both  \toolnameBH{}  and OpenOrd scale linearly with threads (cores), but  \toolnameBH{} is $\sim4\times$ faster than OpenOrd on all thread counts. 
ForceAtlas2BH does not scale linearly, and it is generally not faster than OpenOrd.
Fig.~\ref{fig:selfcheckout2}(c) shows the \commentKhaled{weak scalability where we use increasingly larger graphs varying number of threads. Fig.~\ref{fig:selfcheckout2}(c) starts with a graph of $2^{16}$ vertices running on 2 threads. Then, we increasingly double the problem size as well as computing resource, run experiments and report the runtime. Among all three algorithms, we observe that \toolnameBH{}'s work distribution remains balanced (i.e., curve remains horizontal). Hence, \toolnameBH{} shows superior weak scaling performance compared to other two algorithms.}

{\bf Memory usage.} We measured the memory consumption of all algorithms using the {\myfont memory-profiler} python package.
For the lxOSM graph, \toolname{}, \toolnameBH{}, ForceAtlas2, ForceAtlas2BH, OpenOrdG, and OpenOrd consumed maximum memory of 16.36MB, 30.34MB, 676.38MB, 1348.38MB, 1265.32MB and 14.12MB, respectively. 
Gephi's ForceAtlas2, ForceAtlas2BH and OpenOrdG consumed higher memory than others possibly because of the high memory requirements of Gephi itself.
In fact, ForceAtlas2BH failed to generate layouts for two large graphs due to its high memory consumption.
In our algorithm, Barnes-Hut force approximation requires additional space to store the quad-tree data structure, and hence it consumed more memory than the exact force computation. 
Overall, \toolnameBH{} can compute layouts faster without consuming significant memory.

\begin{table}[!tb]
\caption{Runtime for different methods. Lower value represents better result. BL, BLBH, FA2, FA2BH, OOG and OO represent \toolname{}, \toolnameBH{}, ForceAtlas2, ForceAtlas2BH, OpenOrdG and OpenOrd, respectively. Better results are shown in bold font.}
\vspace{-4pt}
\centering
\begin{tabular}{|p{1.2cm}|p{0.63cm}|p{0.6cm}|p{0.65cm}|p{0.7cm}|c|p{0.8cm}|}
\hline
\multirow{2}{*}{\textbf{Graph}} & \multicolumn{6}{c|}{\textbf{Running time (seconds)}} \\ \cline{2-7} 
                                & BL   & BLBH   & FA2   & FA2BH   & OOG   & OO  \\ \hline
Powergrid	    &   \textbf{0.95}	&   0.98	&   20.73	&   3.04	&   1.67 & 1.43\\ \hline
add32	&   1.03	&   \textbf{1.02}	&   20.75	&   3.15	&   1.71 &   1.45 \\ \hline
ba\_network	&   1.42	&   \textbf{1.22}	&   25.47	&   3.61	&   2.18 & 1.78 \\ \hline
3elt\_dual	&   2.92	&   \textbf{1.54}	&   54.6	&   4.87	&   3.17 & 2.67 \\ \hline
PGP	&   4	&   \textbf{1.83}	&   68.67	&   6.26	&   4.18 & 3.26\\ \hline

pkustk02	&   4.23	&   \textbf{1.98}	&   76.69	&   10.72	&   10.32 & 3.83 \\ \hline
fe\_4elt2	&   4.33	&   \textbf{1.86}	&   75.4	&   6.34	&   4.21 & 3.35 \\ \hline
bodyy6	&   12.3	&  \textbf{2.95}	&   143.08	&   9.69	&   6.75 & 5.85 \\ \hline
pkustk01	&   16.23	&   \textbf{3.43}	&   242.72	&   18.49	&   15.33 & 7.25 \\ \hline
OPF\_6000	&   29.75	&   \textbf{4.69}	&   266.96	&   18.31	&   12.45 & 8.95\\ \hline
finance256	&   45.49	&   \textbf{5.4}    &   412.6	&   20.37	&   14.69 & 11.18\\ \hline
finan512	&   185.8	&   \textbf{10.53}	&   1408	&   46.98	&   31.74 & 22.57\\ \hline
lxOSM	&   -	&   \textbf{21.59}	&   -	&  72.30 	&  -  & 33.62 \\ \hline
comYout.	&   -	&   \textbf{174.5}	&   -	&   1504.7	& -    & 1670.52 \\ \hline
Flan\_1565	&  -	&   \textbf{224.3}	&   -	&   -	&   - & 609.56 \\ \hline
\end{tabular}
\label{tab:runningtime}
\vspace{-0.4cm}
\end{table}

\begin{table*}[!t]
\centering
\caption{Layouts of different graphs generated by different algorithms. BatchLayout convergence criteria was set to 1E-6. Number of iterations and running time are shown in supplementary file, Table S2.}
\label{tab:convergedlayouts}

\begin{tabular}{|c|p{2.5cm}|p{2.6cm}|p{2.6cm}|p{2.6cm}|p{2.5cm}|}
\hline
\textbf{Graph}   & \textbf{BatchLayout} & \textbf{BatchLayoutBH} & \textbf{ForceAtlas2} & \textbf{ForceAtlas2BH} & \textbf{OpenOrd} \\ \hline
        \textbf{Powergrid} & \multicolumn{5}{|c|}{\includegraphics[height=1.8cm,width=0.8\linewidth]{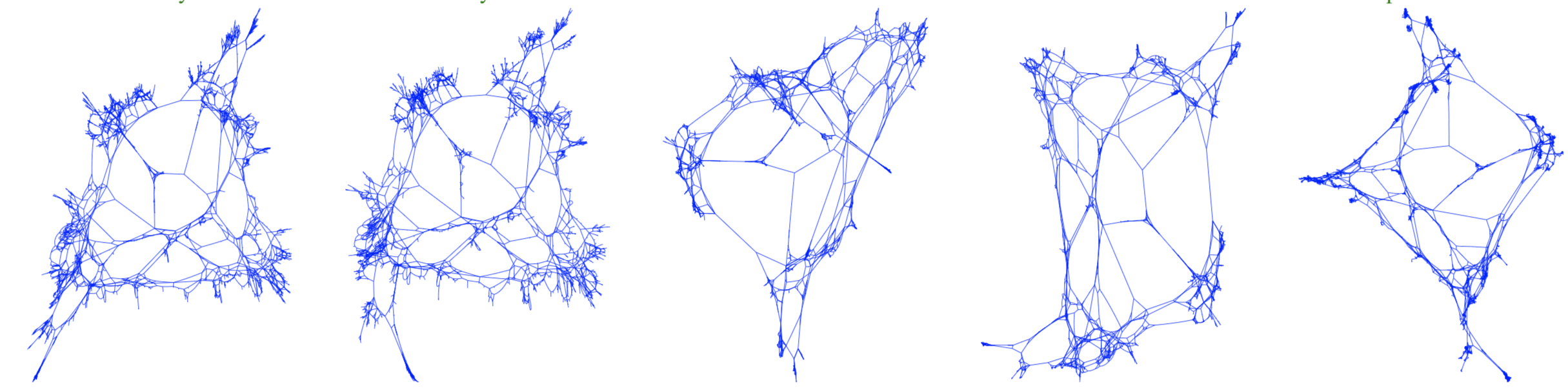}}                                                                                                                \\ \hline

\textbf{ba\_network}        & \multicolumn{5}{c|}{\includegraphics[height=1.8cm,width=0.8\linewidth]{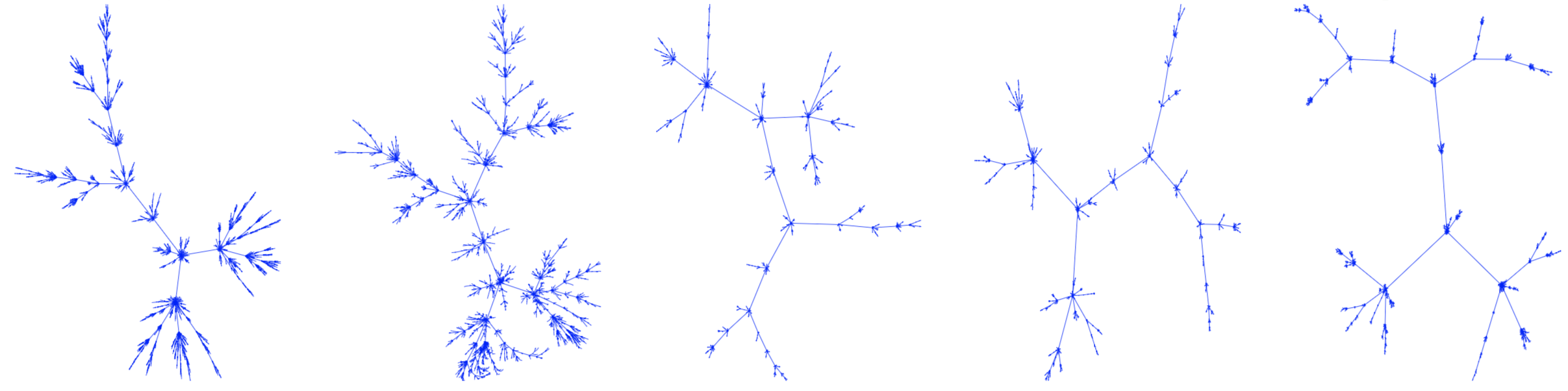}}                                                                                                                 \\ \hline
\textbf{3elt\_dual}        & \multicolumn{5}{c|}{\includegraphics[height=1.8cm,width=0.8\linewidth]{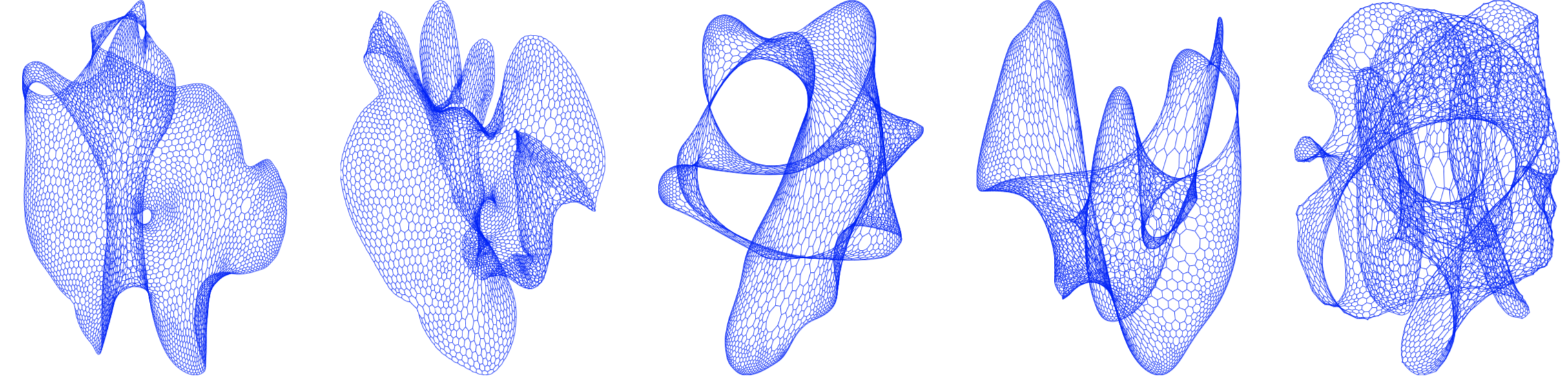} }                                                                                                                                                        \\ \hline
\textbf{pkustk02}        & \multicolumn{5}{c|}{\includegraphics[height=1.8cm,width=0.8\linewidth]{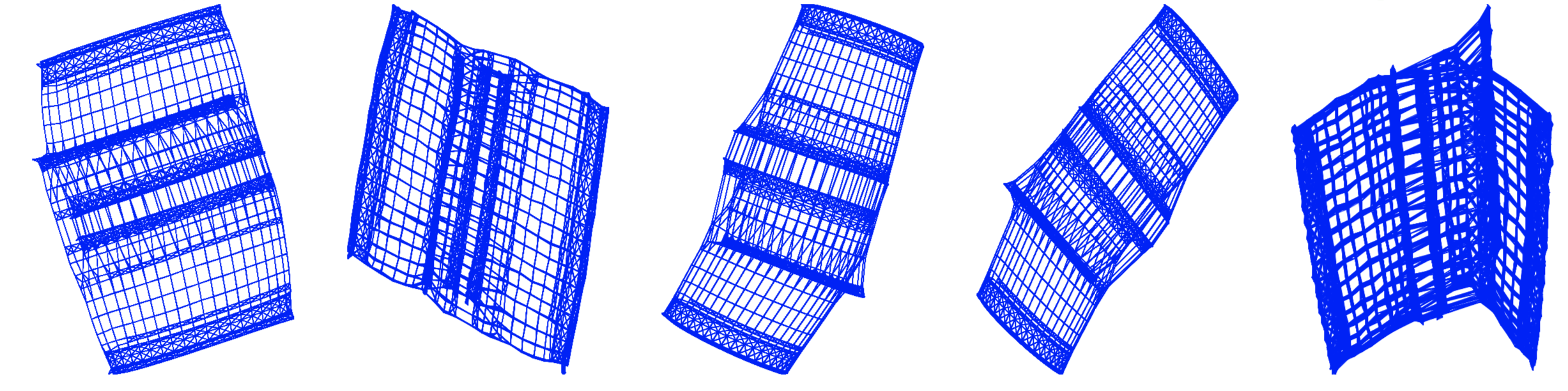}}                                                                                                                 \\ \hline
\textbf{fe\_4elt2}        & \multicolumn{5}{c|}{\includegraphics[height=1.8cm,width=0.8\linewidth]{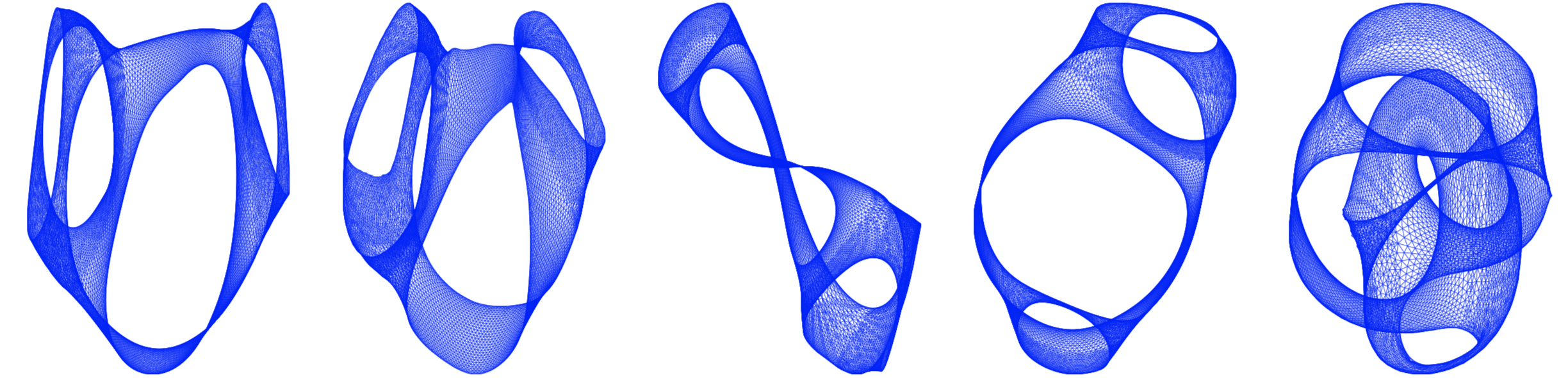}}                                                                                                                 \\ \hline
\textbf{bodyy6}        & \multicolumn{5}{c|}{\includegraphics[height=1.8cm,width=0.8\linewidth]{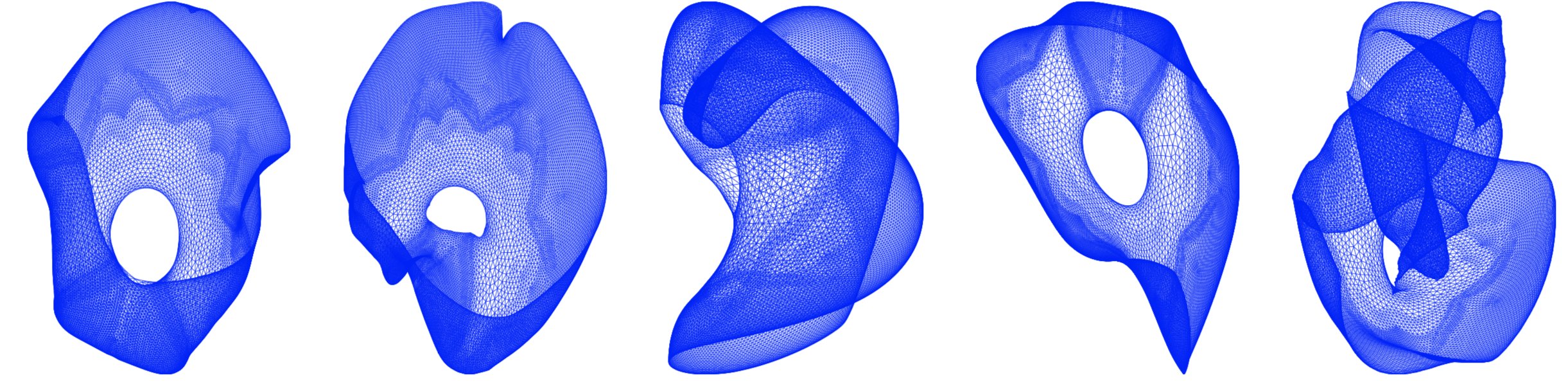}}                                                                                                                 \\ \hline

\textbf{pkustk01}        & \multicolumn{5}{c|}{\includegraphics[height=1.8cm,width=0.8\linewidth]{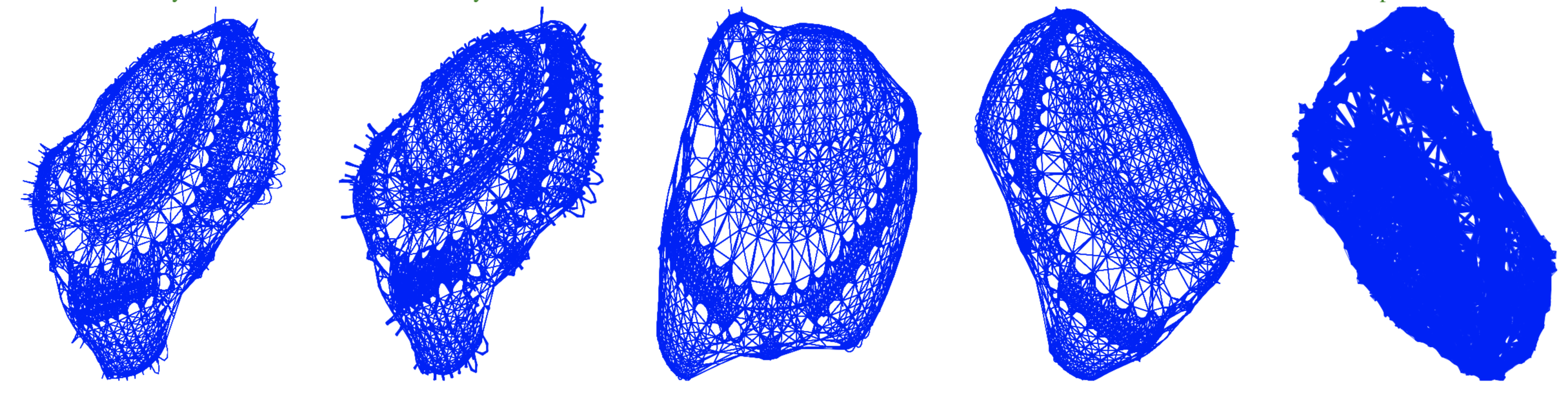}}  \\ \hline

\textbf{finan256}        & \multicolumn{5}{c|}{\includegraphics[height=1.8cm,width=0.8\linewidth]{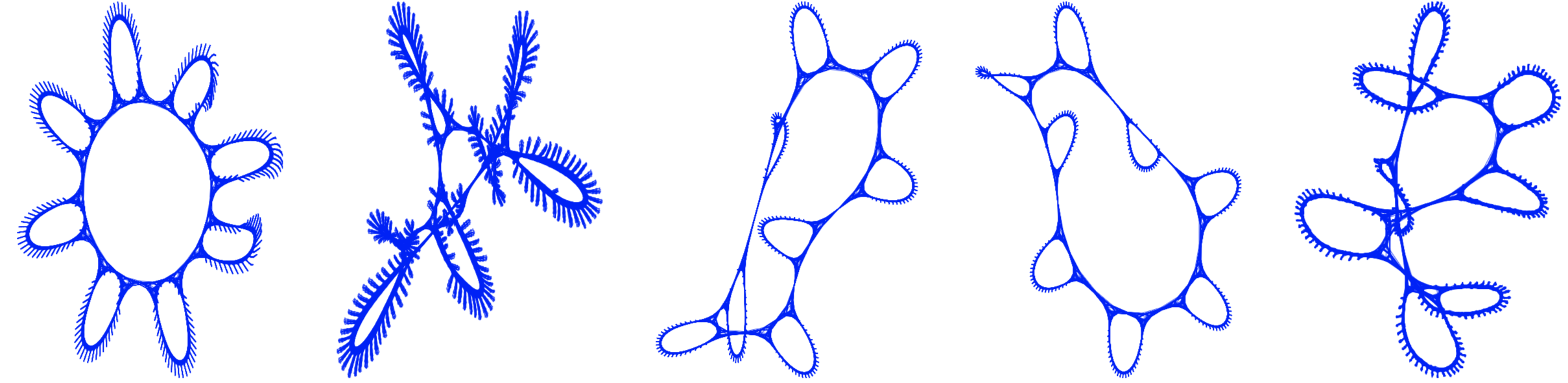}}  \\ \hline
\end{tabular}
\vspace{-0.2cm}
\end{table*}

\subsection{Visualization}
We now visualize the layouts generated by all algorithms considered in the paper. 
At first, we run \toolname{} until convergence (when the energy difference between successive iterations becomes less than $10^{-6}$).
Then, all other algorithms including \toolnameBH{} were run for the same number of iterations that \toolname{} took to converge.
The visualization from these layouts are shown in Table~\ref{tab:convergedlayouts}.
The runtime and converged energy are provided  in supplementary Table S2, which shows that \toolnameBH{} runs much faster than other tools.
Table~\ref{tab:convergedlayouts} demonstrates that \toolname{} and \toolnameBH{} produce \commentKhaled{readable} layouts that are comparable or better than ForceAtlas2 and ForceAtlas2BH, respectively.
Generally, \toolnameBH{} and ForceAtlas2BH layouts are much better than OpenOrd.
Gephi's OpenOrd generates inferior layouts; hence, we did not show them in Table~\ref{tab:convergedlayouts}.
Note that \toolname{} uses Fruchterman and Reingold as its default energy model, which is different from the default model in ForceAtlas2. 
Hence, \toolname{}'s superior layouts for some graphs (e.g., finan256) are not surprising because the FR model usually generates better layouts as was also reported in the ForceAtlas2 paper~\cite{jacomy2014forceatlas2}.
ForceAtlas2 does not use FR as its default model possibly because of its higher computational cost (also reported in ~\cite{jacomy2014forceatlas2}).
In this paper, we show that the FR algorithm runs much faster in our \toolname{} framework and can generate better layouts for some graphs. 

In the scalability and runtime experiments, we showed results with a fixed 500 iterations. 
Supplementary Table S1 shows the visualization from all algorithms after 500 iterations.
In many cases, layouts after 500 iterations are close to the layouts at convergence. 
Hence, many algorithms including \toolname{} provide ``number of iterations" as an option to users.   

Both \toolname{} and ForceAtlas2 provide many options beside default options used in our experiments. 
It is possible that tuning these parameters will generate better-quality layouts for some graphs.
However, we stick to default parameters in this paper for simplicity and fairness. 
OpenOrd may generate different layouts for the same graph when different numbers of threads are used.
By contrast, \toolname{} and \toolnameBH{} generate the same layout for a given graph irrespective of number of threads.

\begin{figure}[!t]
\vspace{-0.1cm}
    \centering
    \includegraphics[width=0.48\linewidth]{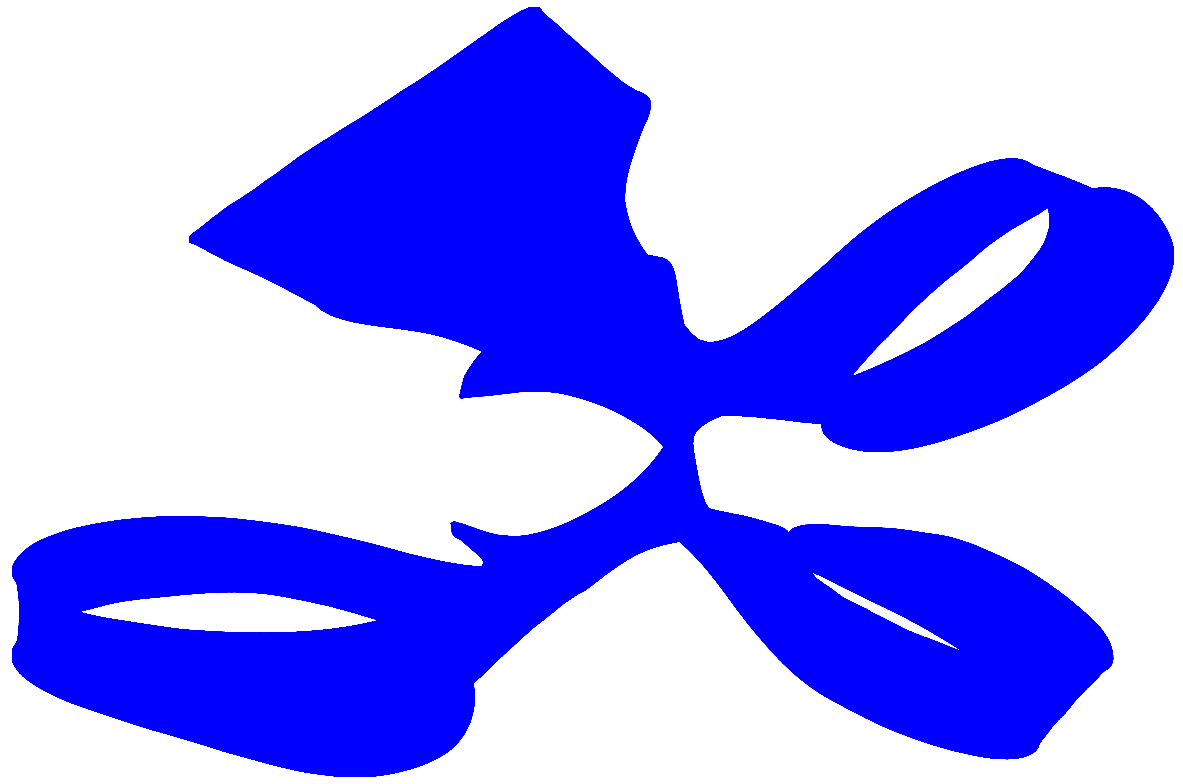}
    \includegraphics[width=0.48\linewidth]{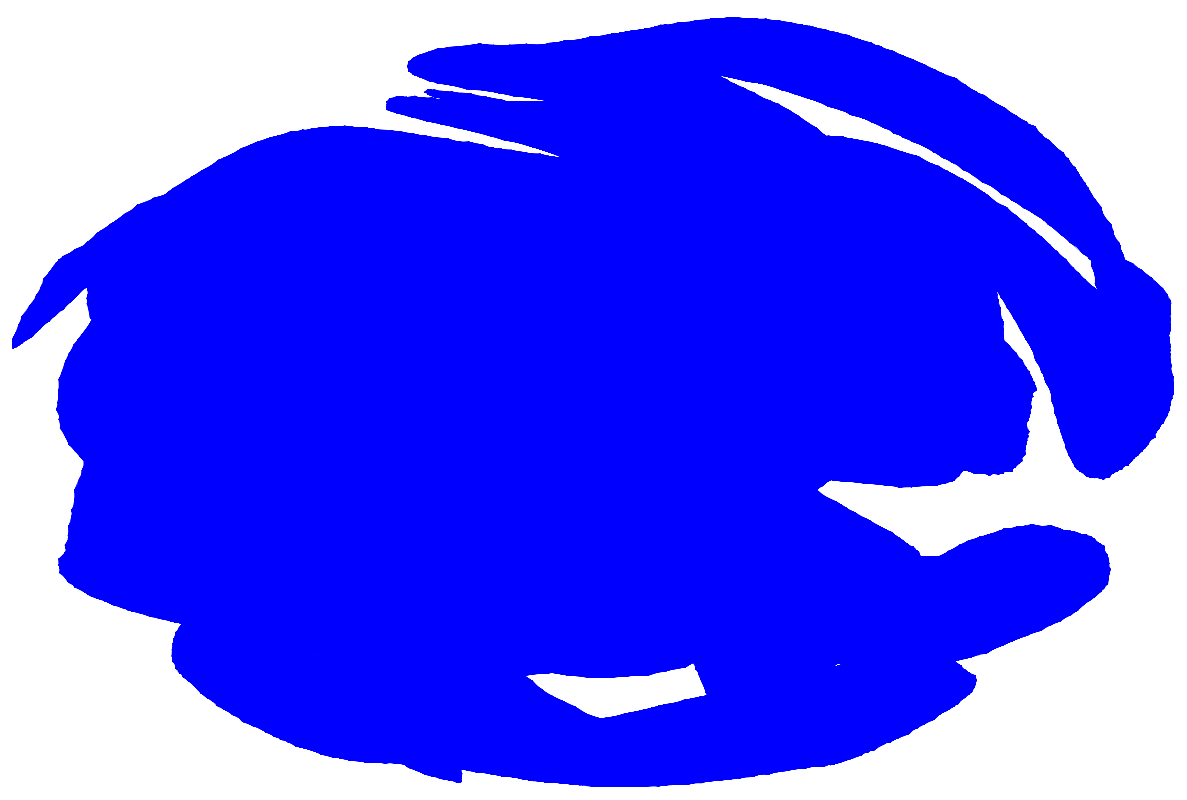}
    \caption{Layouts of \emph{Flan\_1565} graph (1.56M vertices and 114.2M edges) generated by \toolnameBH{} (left) and
    OpenOrd (right).
    }
    \vspace{-0.4cm}
    \label{fig:verybiggraph}
\end{figure}

Finally, Fig.~\ref{fig:verybiggraph} shows the layout of \emph{Flan\_1565}, the largest graph in our dataset.
We show layouts from \toolnameBH{} and OpenOrd after running them for 5000 iterations (ForceAtlas2BH went out of memory for this graph).
For this graph,  OpenOrd's layout is not good enough even though we allow both algorithms to run for 5000 thousands of iterations. 
Furthermore, OpenOrd took 1 hour and 39 minutes whereas \toolnameBH{} took 49 minutes to finish 5000 iterations. 
This clearly demonstrates the effectiveness of \toolnameBH{} to visualize graph with hundreds of millions of edges.
However, generating layouts in half an hour may not be good enough for certain applications.
We plan to address this by implementing  \toolnameBH{} for distributed systems.

\begin{table*}[!t]
\caption{Comparison of Stress (ST) and Neighborhood preservation (NP) measures among all tools. For ST, a lower value means a better result and for NP, a higher value represents a better result. Better results are shown in bold font.}
\vspace{-4pt}
\centering
\begin{tabular}{|c|c|c|c|c|c|l|c|c|c|c|c|c|}
\hline
\multirow{2}{*}{\textbf{Graph}} & \multicolumn{6}{c|}{\textbf{ST}}                            & \multicolumn{6}{c|}{\textbf{NP}}            \\ \cline{2-13} 
                                & BL       & BLBH     & FA2       & FA2BH & OO & OOGH & BL & BLBH & FA2 & FA2BH & OOG & OO \\ \hline

Powergrid	&	\textbf{1.3E+6}	&	1.6E+6	&	2.4E+6	&	2.2E+6	&	2.5E+6	&	2.4E+6	&		0.39	&	0.266	&	0.403	&	\textbf{0.408}	&	0.324	&	0.336 \\ \hline

add32	&	2.2E+6	&	\textbf{1.9E+6}	&	5.3E+6	&	5.4E+6	&	6.2E+6	&	2.3E+6	&			0.453	&	0.347	&	\textbf{0.538}	&	0.523	&	0.362	&	0.409 \\ \hline

ba\_network	&	2.8E+6	&	\textbf{2.5E+6}	&	3.4E+6	&	3.4E+6	&	3.9E+6	&	2.9E+6	&			0.357	&	0.295	&	0.457	&	0.452	&	0.384	&	\textbf{0.465} \\ \hline

3elt\_dual	&	\textbf{5.6E+6}	&	6.6E+6	&	9.2E+6	&	8.0E+6	&	1.4E+7	&	1.1E+7	&			\textbf{0.38}	&	0.271	&	0.253	&	0.334	&	0.171	&	0.174 \\ \hline

PGP	&	\textbf{8.2E+6}	&	1.2E+7	&	9.9E+6	&	9.8E+6	&	1.1E+7	&	8.8E+6	&			0.18	&	0.119	&	0.24	&	0.243	&	\textbf{0.266}	&	0.231 \\ \hline

pkustk02	&	\textbf{3.7E+6}	&	4.4E+6	&	1.0E+7	&	1.1E+7	&	1.9E+7	&	8.1E+6	&			\textbf{0.649}	&	0.583	&	0.549	&	0.544	&	0.38	&	0.513 \\ \hline

fe\_4elt2	&	\textbf{1.0E+7}	&	1.7E+7	&	1.2E+7	&	1.5E+7	&	2.1E+7	&	1.8E+7	&			\textbf{0.36}	&	0.25	&	0.319	&	0.289	&	0.222	&	0.207 \\ \hline

bodyy6	&	\textbf{3.1E+7}	&	4.1E+7	&	3.4E+7	&	4.3E+7	&	6.1E+7	&	5.2E+7	&			0.213	&	0.205	&	\textbf{0.278}	&	0.251	&	0.217	&	0.124 \\ \hline

pkustk01	&	\textbf{1.7E+7}	&	1.9E+7	&	1.8E+7	&	2.0E+7	&	4.1E+7	&	3.1E+7	&			0.513	&	0.398	&	\textbf{0.528}	&	0.523	&	0.373	&	0.396 \\ \hline

\end{tabular}
\label{tab:measures_st_np}
\vspace{-0.2cm}
\end{table*}

\begin{table}[!t]
\caption{\commentKhaled{Comparison of Edge uniformity (EU) measures among all tools. For this measure, lower value means better result.} Better result of each graph is shown in bold font.}
\vspace{-4pt}
\centering
\begin{tabular}{|p{1.3cm}|c|p{0.7cm}|c|c|c|c|}
\hline
\multirow{2}{*}{\textbf{Graph}} &  \multicolumn{6}{c|}{\textbf{EU}}            \\ \cline{2-7} 
                                & BL & BLBH & FA2 & FA2BH & OOG & OO \\ \hline

Powergrid	&     0.83	&   \textbf{0.5}	&   1.46	&   1.33	&   1.93	&   1.18 \\ \hline

add32	&   1.38	&   \textbf{1.04}	&   1.30	&   1.26	&   1.86	&   1.76 \\ \hline

ba\_network	&     1.07	&   \textbf{0.58}	&   3.23	&   3.30	&   3.01	&   2.93 \\ \hline

3elt\_dual	&   0.39	&   \textbf{0.39}	&   0.59	&   0.63	&   1.37	&   0.53 \\ \hline

PGP	&  0.81	&   \textbf{0.65}	&   1.43	&   1.42	&   2.44	&   1.43 \\ \hline

pkustk02		&   0.73	&   \textbf{0.67}	&   1.09	&   1.12	&   1.93	&   0.88 \\ \hline

fe\_4elt2	&   \textbf{0.39}	&   0.53	&   0.53	&   0.51	&   1.55	&   0.48 \\ \hline

bodyy6	&   0.76	&   0.80	&   \textbf{0.49}	&   0.51	&   1.64	&   0.79 \\ \hline

pkustk01	&   	0.83	&   \textbf{0.70}	&   1.01	&   1.02	&   1.62	&  0.93 \\ \hline
\end{tabular}
\vspace{-0.5cm}
\label{tab:measures_ec_eu}
\end{table}

\subsection{Comparison of Aesthetic Metrics}
We quantitatively measure the quality of graph layouts using standard aesthetic metrics discussed in Section~\ref{sec:aesthetic_measure}.
Similar to the runtime analysis, we measure aesthetic quality of layouts after running each algorithm for 500 iterations and report the numbers in Tables~\ref{tab:measures_st_np} and \ref{tab:measures_ec_eu}.
\commentKhaled{There is no clear winner according to all metrics, which is expected since different algorithms optimize different energy functions. Table \ref{tab:measures_st_np} shows ST (lower is better) and NP measures (higher is better), where \toolname{} and \toolnameBH{} are better than other tools for ST measure and competitive with their peers for NP measure. Table \ref{tab:measures_ec_eu} shows EU (lower is better), where we observe that \toolnameBH{} is winner for most of the graph instances. Overall, \toolname{} and \toolnameBH{} perform better according to ST and EU measures and competitive according to NP measure.}

\section{Related Work}
\label{sec:relatedworks}
Graph drawing is a well studied problem in the literature. 
In addition to the spring-electrical approach, there are other models such as the spectral method \cite{koren2003drawing}, and high-dimensional embedding \cite{harel2002graph}. 
Recently, Kruiger et al.~\cite{kruiger2017graph} introduced tsNET based on stochastic neighbor embedding technique. 
Due to space limitation, we restrict our discussion to force-directed approaches.

Kamada and Kawai~ \cite{kamada1989algorithm} proposed a spring model where optimal drawing is obtained by minimizing stress.
In their model, stress is represented by the difference between geometric distance and graph-theoretic shortest path distance. 
The computational complexity of this model is high, and quad-tree based force approximation can not be applied to it~\cite{hu2005efficient}. 
Other graph drawing algorithms that optimize a stress function also produce readable layouts~\cite{gansner2012maxent,meyerhenke2017drawing}, but they are generally prohibitively expensive for large graphs.

Fruchterman and Reigngold (FR) proposed a spring-electrical model where an energy function is defined in terms of \emph{attractive} and \emph{repulsive} forces \cite{fruchterman1991graph}. 
This model is computationally faster than the spring model, and quad-tree based force approximation can also be applied. 
Yifan Hu introduced a multi-level version of force-directed algorithm~\cite{hu2005efficient}. In his approach, repulsive force can be approximated by a quad-tree data structure, resulting in one of the fastest sequential algorithms. \commentKhaled{Other efficient multi-level algorithm includes $FM^3$ \cite{hachul2004drawing} which was later integrated in OGDF library \cite{chimani2013open}. 
Interestingly, Andreas Noack introduced the LinLog energy model which is very effective in capturing clusters in the graph layout~\cite{noack2003energy}.

OpenOrd~\cite{martin2011openord} by Martin et al. is a parallel multi-level algorithm following the force-directed model.}
OpenOrd is reasonably fast and generate layouts of large-scale graphs. 
Finally, ForceAtlas2 by Jacomy et al. introduced a more general framework for graph layout combining various features (repulsive force approximation by Barnes-Hut approach, LinLog mode, etc.)~\cite{jacomy2014forceatlas2}. 
ForceAtlas2 is known to generate continuous layouts as a user can stop the program anytime with valid (possibly unoptimized) layouts.
Our work is influenced by ForceAtlas2 and covers many features available in ForceAtlas2.
The ForceAtlas2 paper~\cite{jacomy2014forceatlas2} demonstrated that FR model can generate \commentKhaled{readable} layouts but it can be slower than other energy models. 
In this paper, we addressed this challenge with a multi-threaded cache-efficient algorithm that is both faster and generates \commentKhaled{readable} layouts.

Recently, Zheng et al. used a sequential SGD approach for graph drawing, which optimizes stress to generate the layout of a graph~\cite{sgd28419285}. 
However, their software is very slow as it considers graph theoretic distances in the optimization function. 
Both ForceAtlas2 and OpenOrd can run in parallel and are available in Gephi's toolkit~\cite{bastian2009gephi}. 
Force-directed algorithm has also been implemented in \commentKhaled{distributed platforms~\cite{arleo2017large,arleo2018distributed}} as well as for GPUs~\cite{brinkmann2017exploiting}.
These tools  cannot be run on any server because they rely on special hardware. 
We did not compare with these implementations because our objective in this paper is to develop a general-purpose and fast algorithm for multicore servers.

\vspace{-0.2cm}
\section{Discussions and Conclusions}
In this paper, we present a parallel force-directed algorithm \toolname{} and its Barnes-Hut approximation  \toolnameBH{} for generating 2D layouts of graphs. 
The presented algorithms are highly scalable, robust with respect to diverse classes of graphs, can be run on any multicore computer, and runs faster than other state-of-the-art algorithms. 
Aside from carefully chosen default hyper-parameters (e.g., initialization, minibatch size, energy model, learning rate, convergence conditions), \toolname{} provides flexibility to let users choose hyper-parameters that are suitable for the graph.
In terms of flexibility, \toolname{} is comparable to the flexibility of ForceAtlas2 and can be integrated with any visualization tool. 

The high performance of \toolname{} comes from two important optimizations that we made. 
First, \toolname{} exposes more parallel work to keep many processors busy. 
More parallel work comes from our minibatch scheme, which is also a widely-used technique in training Deep Neural Networks. 
Second, we incorporate cache-blocking technique commonly used to optimize linear algebra operations.
Hence, this paper unifies two proven techniques from machine learning and linear algebra and delivers a high-performance graph layout algorithm.
As a result, \toolname{} is highly scalable and runs significantly faster than other state-of-the-art tools. 



\toolname{} generates graph layouts without sacrificing their aesthetic qualities.
We visually and analytically verified the quality of layouts for different classes of graphs covering grid networks, small world networks, scale-free networks, etc.
In all cases, \toolname{} generates good layouts that are similar or better than its peers.



\toolname{} is implemented as an open-source software with detailed documentation. 
This software depends on standard C++ and OpenMP libraries and is portable to most computers.
We use simple Python scripts to visualize layouts generated by our software. 
Hence, we believe that \toolname{} will benefit many users in visually analyzing complex networks from diverse scientific domains.

While this paper only focuses on \commentKhaled{parallel} force-directed algorithms, other classes of algorithm might generate better layouts than the algorithms considered in this paper.
For example, tsNET can generate better quality layouts when it converges. 
However, such tools are often prohibitively expensive for large-scale graphs.
For example, tsNET took 16 hours to generate a layout for \emph{OPF\_6000}, whereas \toolname{} generates a comparable layout in just few seconds.
In fact, users can set a very low threshold in \toolname{}, decrease the batch size and increase number of iterations to get superior layouts from our software (e.g., see supplementary Table S4). \commentKhaled{We also compared our results with $FM^3$ available in OGDF library and found that \toolnameBH{} is always faster than $FM^3$ for non-engineered settings in OGDF library (see supplementary Table S9 for runtime details). For two large graphs of Table \ref{tab:datasets}, \toolnameBH{} is approximately $5\times$ faster than $FM^3$.}


This paper only considered shared-memory parallel algorithms since multicore computers and servers are prevalent in scientific community.
However, \toolname{} can be easily implemented for GPUs and distributed-memory systems. 
\commentKhaled{A distributed algorithm will require larger minibatches if only ``data parallelism" is used. 
However, if we also use graph partitioning (which is equivalent to model parallelism in deep learning), we can have enough parallelism for large-scale distributed systems.} 
Expanding \toolname{} for other high-performance architectures and improving the aesthetic quality for massive graphs remain our future work.



\section*{Acknowledgements}
\commentKhaled{We would like to thank Stephen Kobourov, Katy Borner, Iqbal Hossain, Felice De Luca and Bruce Herr for helpful discussions and comments on measures. 
Funding for this work was provided by the Indiana University Grand Challenge Precision Health Initiative.}

\vspace{-0.15cm}
\nocite{*}

\end{document}


\setcounter{page}{1}
\title{BatchLayout: A Batch-Parallel Force-Directed Graph Layout Algorithm in Shared Memory\\ (supplementary file)}
\author{Md. Khaledur Rahman, Majedul Haque Sujon and Ariful Azad}
\maketitle
\section{Online Repository}
\toolname{} is publicly available at \url{https://github.com/khaled-rahman/BatchLayout}. Benchmark datasets, results and well-documented instructions also available in the repository. Note that we have also shared other tools and instructions about how to run. For OpenOrd, we have both Gephi's version and author's version.

\section{How to compile \toolname{}}
We will be continuing to improve performance \toolname{} using different optimization techniques such as vectorized operations which will be available at \url{https://github.com/khaled-rahman/BatchLayoutCode}. To compile our source code at \url{https://github.com/khaled-rahman/BatchLayout} without errors, users must have OpenMP version 4.5 and GCC version 4.9. We compiled and run it on linux and macOS. To compile source code, user can type following commads:\newline
\begin{markdown}
```
make clean;
make
```
\end{markdown}

\section{How to run \toolname{}}
Once you successfully compiled source code, type following commad to run \toolname{}:\newline
\begin{markdown}
```
./BatchLayout -input 3elt_dual.mtx -output output/ -iter 600 -batch 256 -threads 32 -algo 2
```
\end{markdown}

Here, parameters have following meaning: 
\newline
\begin{markdown}
```
-input <filename>, full path with file name.
```
\end{markdown}
\newline
\begin{markdown}
```
-output <dir>, output directory where layout is saved.
```
\end{markdown}
\newline
\begin{markdown}
```
-iter <int>, an integer indicates number of iteration. Detaulf:600.
```
\end{markdown}
\newline
\begin{markdown}
```
-batch <int>, an integer indicates size of batch. Default:256.
```
\end{markdown}
\newline
\begin{markdown}
```
-threads <int>, an integer indicates number of threads. Default is max number of threads available in the computing machine.
```
\end{markdown}
\newline
\begin{markdown}
```
-algo <int>, an integer among 0 to 6 which indicates a particular variation of BatchLayout. Default:2. 
```
\end{markdown}

\FloatBarrier
\begin{table*}[htbp]
\centering
\caption{Layouts of different graphs generated by different tools.}
\label{tab:layouts}

\begin{tabular}{|c|p{1.2cm}|p{1.4cm}|p{1.2cm}|p{1.4cm}|p{1.2cm}|p{1.2cm}|}
\hline
\textbf{Graph}   & \textbf{BL} & \textbf{BLBH} & \textbf{FA2} & \textbf{FA2BH} & \textbf{OOG} & \textbf{OO} \\ \hline
        \textbf{Powergrid} & \multicolumn{6}{|c|}{\includegraphics[height=2.05cm,width=0.8\linewidth]{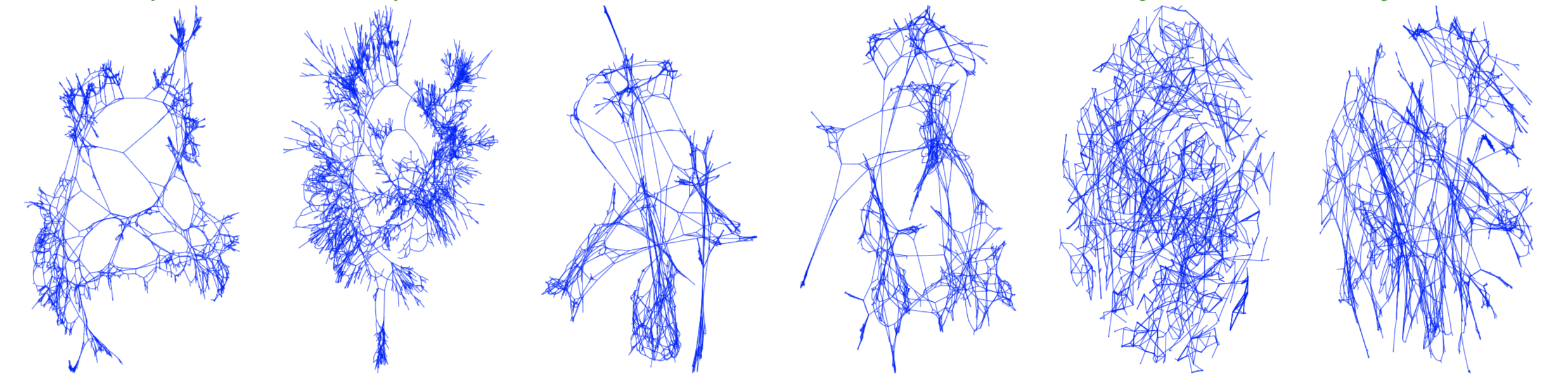}}                                                                                                                \\ \hline
\textbf{add32}   & \multicolumn{6}{c|}{\includegraphics[height=2.05cm,width=0.8\linewidth]{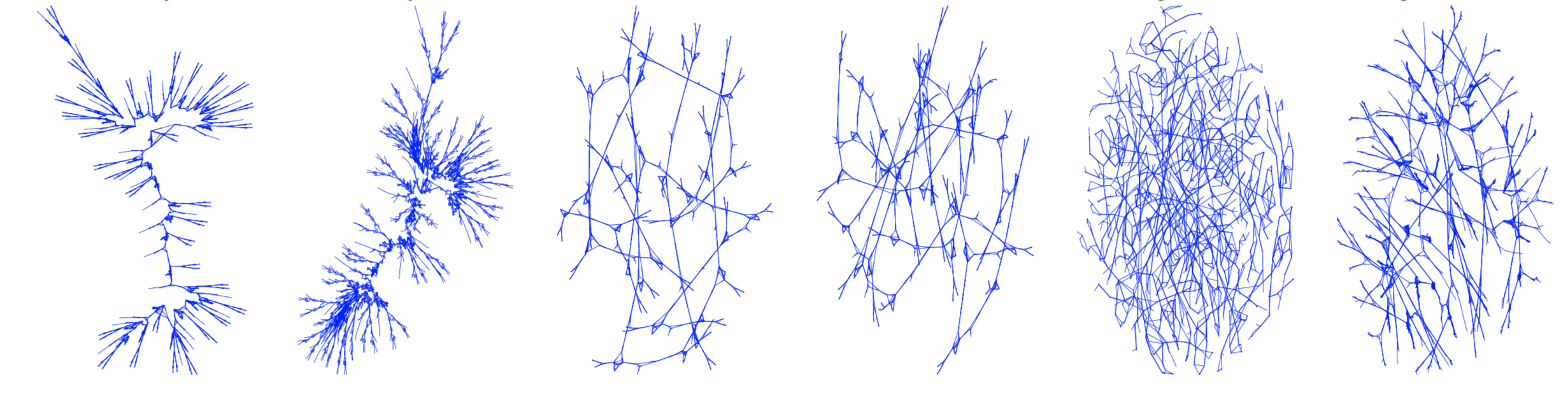}}                                                                                                                 \\ \hline
\textbf{ba\_network}        & \multicolumn{6}{c|}{\includegraphics[height=2.05cm,width=0.8\linewidth]{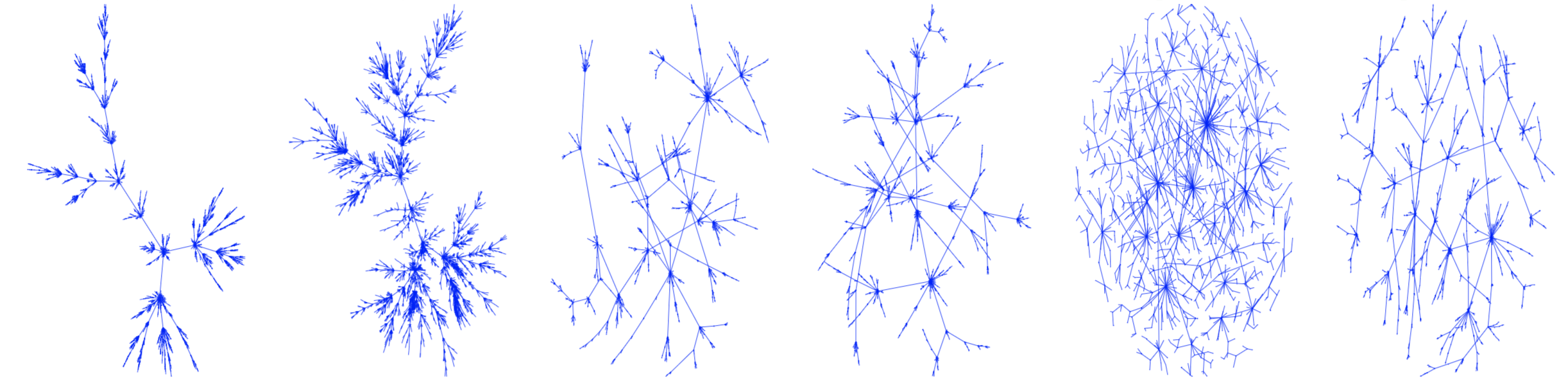}}                                                                                                                 \\ \hline
\textbf{3elt\_dual}        & \multicolumn{6}{c|}{\includegraphics[height=2.05cm,width=0.8\linewidth]{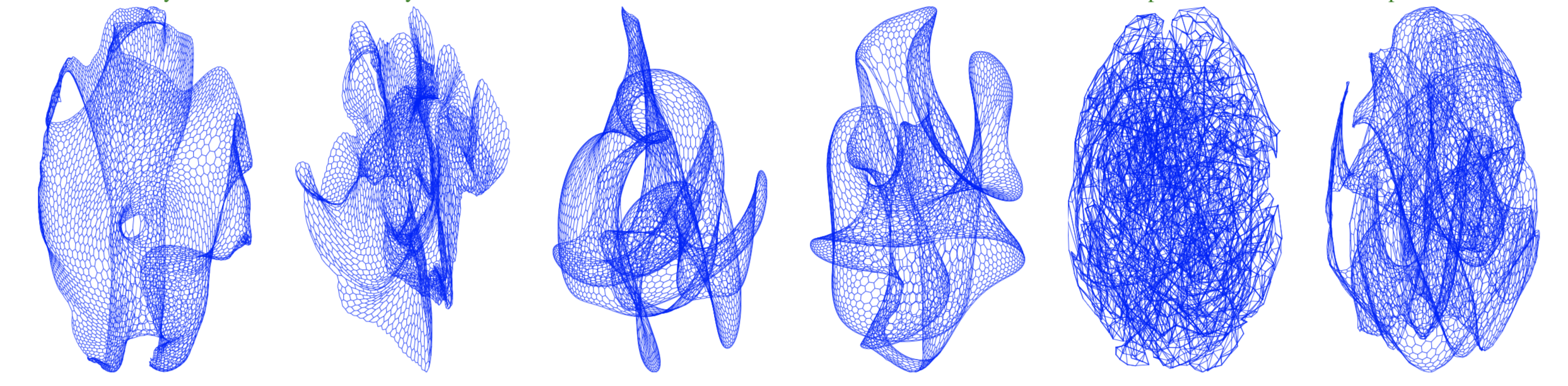}}                                                                                                                 \\ \hline
\textbf{PGPgiant.}        & \multicolumn{6}{c|}{\includegraphics[height=2.05cm,width=0.8\linewidth]{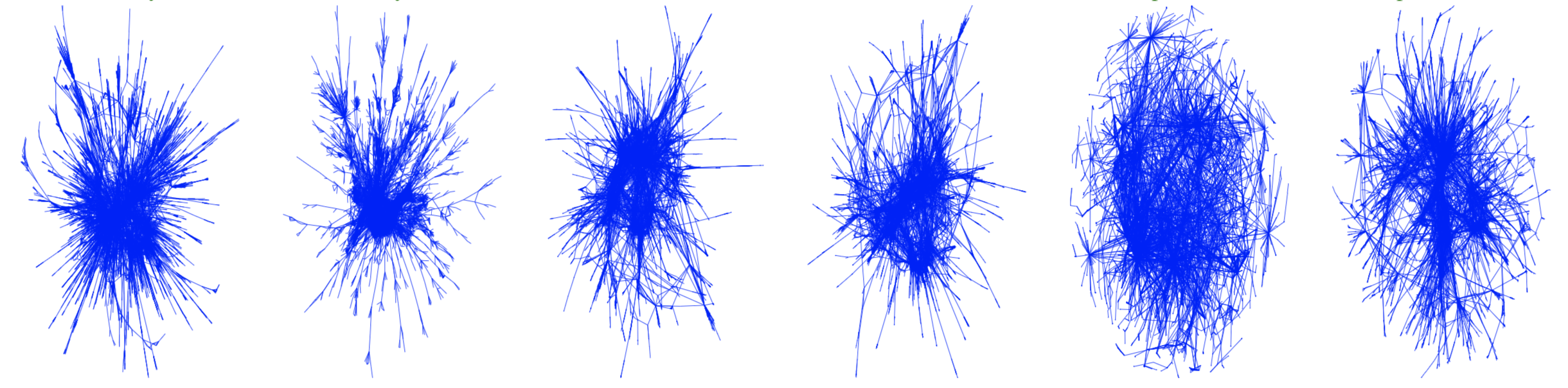}}                                                                                                                 \\ \hline
\textbf{pkustk02}        & \multicolumn{6}{c|}{\includegraphics[height=2.05cm,width=0.8\linewidth]{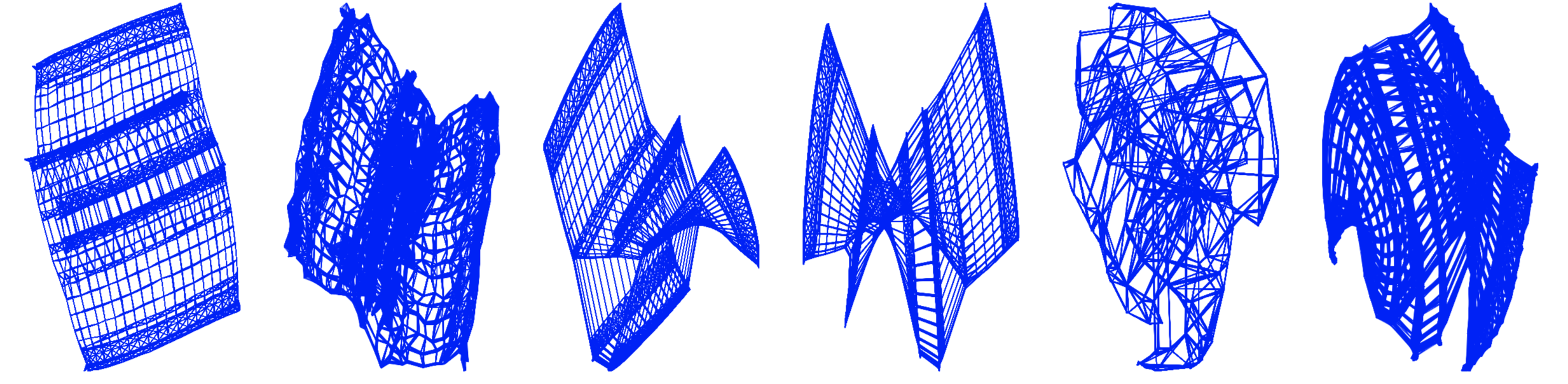}}                                                                                                                 \\ \hline
\textbf{fe\_4elt2}        & \multicolumn{6}{c|}{\includegraphics[height=2.05cm,width=0.8\linewidth]{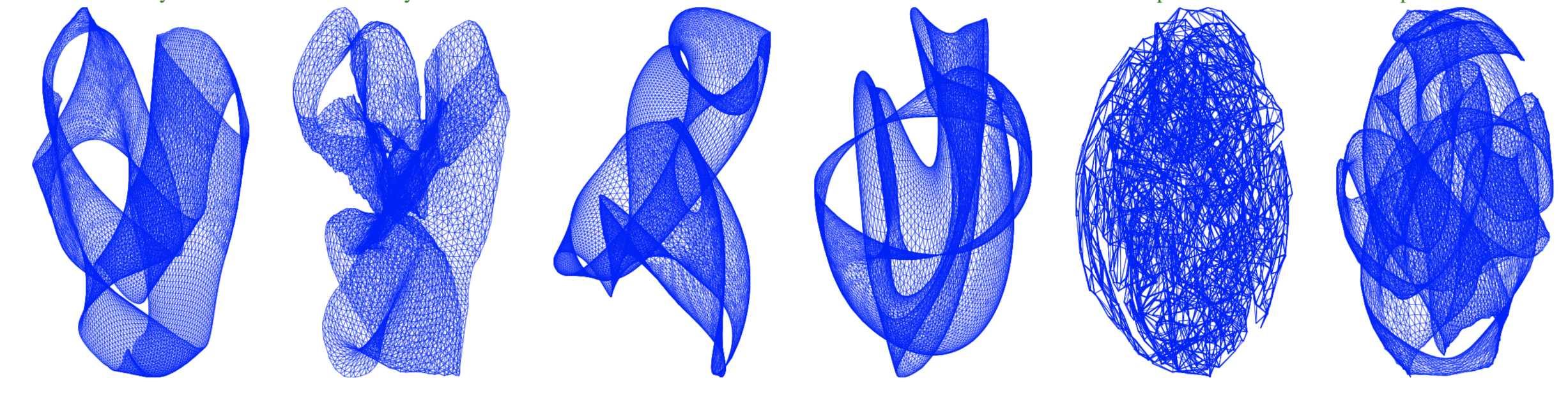}}                                                                                                                 \\ \hline
\textbf{bodyy6}        & \multicolumn{6}{c|}{\includegraphics[height=2.05cm,width=0.8\linewidth]{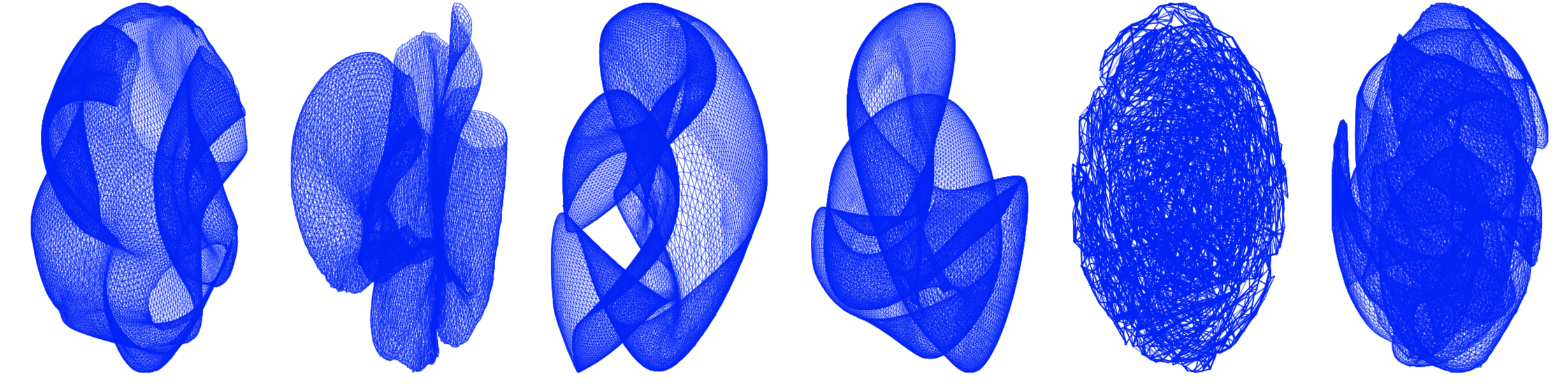}}                                                                                                                 \\ \hline

\textbf{pkustk01}        & \multicolumn{6}{c|}{\includegraphics[height=2.05cm,width=0.8\linewidth]{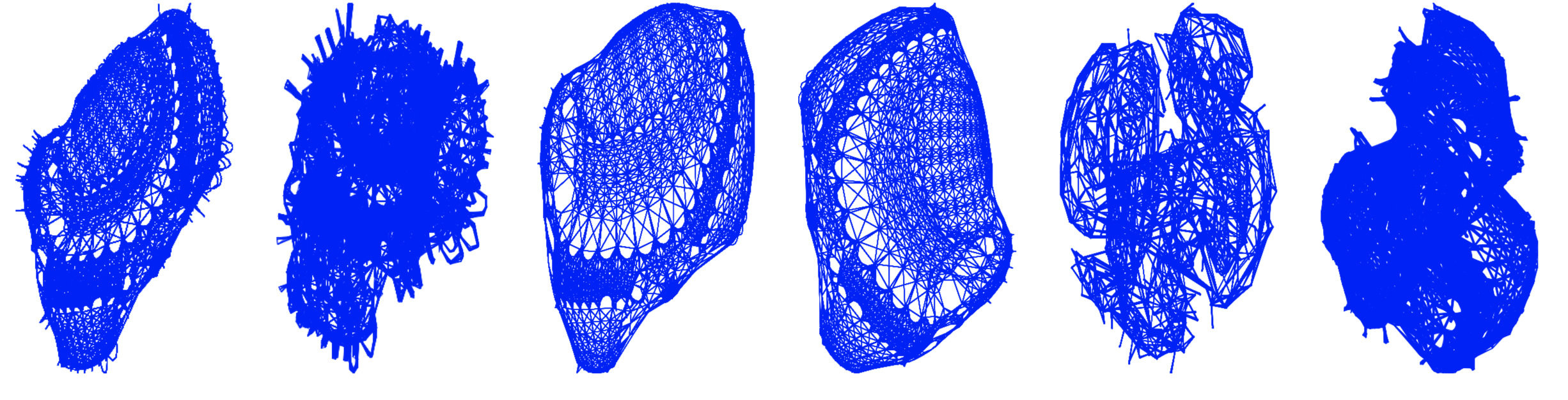}}  \\ \hline

\end{tabular}
\end{table*}

%
\begin{table}[!htb]
\caption{Convergence of BatchLayout for a threshold value of 1E-6 and values of `Iteration' and `Energy' are for converged layouts. Running time in seconds for other tools for corresponding number of iterations that BatchLayout takes to converged.}
\centering
\label{tab:contime}
\begin{tabular}{|c|c|c|
>{\columncolor[HTML]{C0C0C0}}c |
>{\columncolor[HTML]{C0C0C0}}c |
>{\columncolor[HTML]{C0C0C0}}c |
>{\columncolor[HTML]{C0C0C0}}c |
>{\columncolor[HTML]{C0C0C0}}c |}
\hline
\textbf{Graph} & \textbf{Iter.} & \textbf{Energy} & \textbf{BL} & \textbf{FA2} & \textbf{OO} & \textbf{BLBH}  & \textbf{FA2BH} \\ \hline
power\_grid    & 12445                 & 0.036              & \textbf{26.95}   & 553.25            & 31.83                 & 26.99          & 65.32          \\ \hline
ba\_network    & 9216                  & 13.92              & 28.71            & 461.92            & 29                    & \textbf{23.7}  & 59.69          \\ \hline
3elt\_dual     & 1410                  & 1.3E6           & 8.58             & 160.29            & 6.99                  & \textbf{4.21}  & 13.18          \\ \hline
pkustk02       & 1792                  & 2.9E7           & 15.95            & 377.17            & 12.75                 & \textbf{7.52}  & 40.8           \\ \hline
fe\_4elt2      & 3590                  & 298895             & 32.33            & 591.76            & 21.66                 & \textbf{11.99} & 42.92          \\ \hline
bodyy6         & 2042                  & 7.5E6           & 52.2             & 708.63            & 21.51                 & \textbf{11.15} & 39.58          \\ \hline
pkustk01       & 8995                  & 35.853             & 305.49           & 5231.91           & 119.67                & \textbf{72}    & 305.11         \\ \hline
finance256     & 3747                  & 8.5E6           & 354.43           & 4846.48           & 75.19                 & \textbf{40.19} & 152.44         \\ \hline
\end{tabular}
\end{table}

\begin{table}[!htp]
\caption{Distribution of loops for OpenOrd tool used in Table \ref{tab:contime}}
\centering
\begin{tabular}{|c|c|c|c|c|c|c|}
\hline
\textbf{OpenOrd} & \textbf{25\%} & \textbf{25\%} & \textbf{25\%} & \textbf{10\%} & \textbf{15\%} & \textbf{total 100\%} \\ \hline
power\_grid      & 3111          & 3111          & 3111          & 1245          & 1867          & 12445                \\ \hline
ba\_network      & 2304          & 2304          & 2304          & 922           & 1382          & 9216                 \\ \hline
3elt\_dual       & 353           & 353           & 353           & 141           & 210           & 1410                 \\ \hline
pkustk02         & 448           & 448           & 448           & 179           & 269           & 1792                 \\ \hline
fe\_4elt2        & 898           & 898           & 898           & 359           & 537           & 3590                 \\ \hline
bodyy6           & 511           & 511           & 511           & 204           & 305           & 2042                 \\ \hline
pkustk01         & 2249          & 2249          & 2249          & 900           & 1348          & 8995                 \\ \hline
finance256       & 937           & 937           & 937           & 375           & 561           & 3747                 \\ \hline

\end{tabular}
\end{table}

\begin{table}[!htp]
\centering
\caption{Layouts of \emph{3elt\_dual.mtx} graph generated by BatchLayout for different iterations and batches. BS - batch size. Number in each column represents number of iteration and each row represents size of a batch.}
\label{tab:gridlayout}

\begin{tabular}{|p{1.7cm}|p{1.3cm}|p{1.3cm}|p{1.3cm}|p{1.3cm}|p{1.3cm}|p{1.3cm}|}
\hline
\textbf{}   & \textbf{300} & \textbf{600} & \textbf{900} & \textbf{1200} & \textbf{1500} & \textbf{1800} \\ \hline
        \textbf{BS=1} & \multicolumn{6}{|c|}{\includegraphics[height=2.2cm,width=0.8\linewidth]{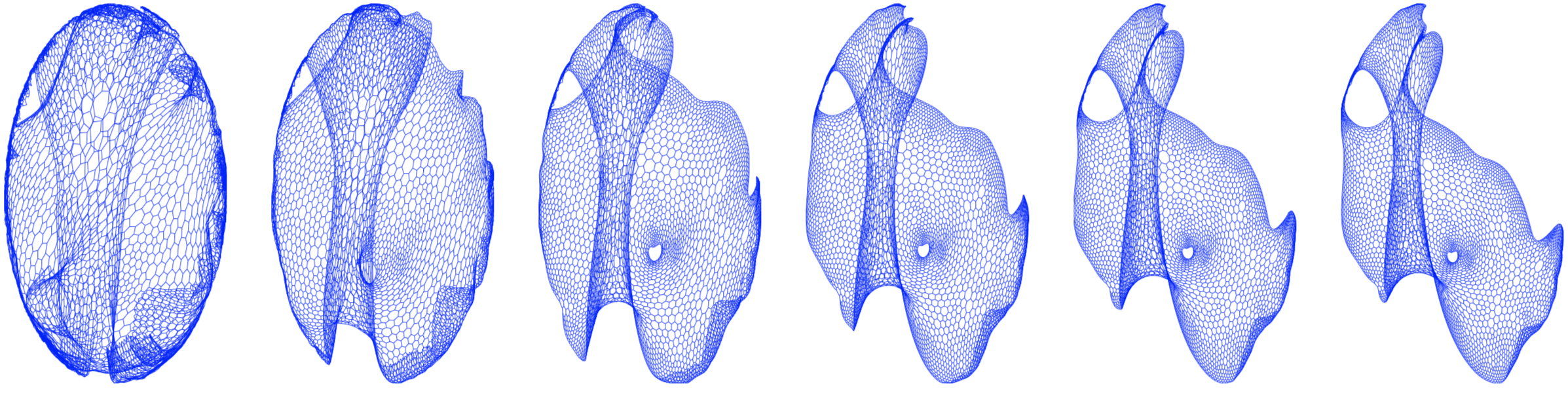}}                                                                                                                \\ \hline

    \textbf{BS=8} & \multicolumn{6}{|c|}{\includegraphics[height=2.2cm,width=0.8\linewidth]{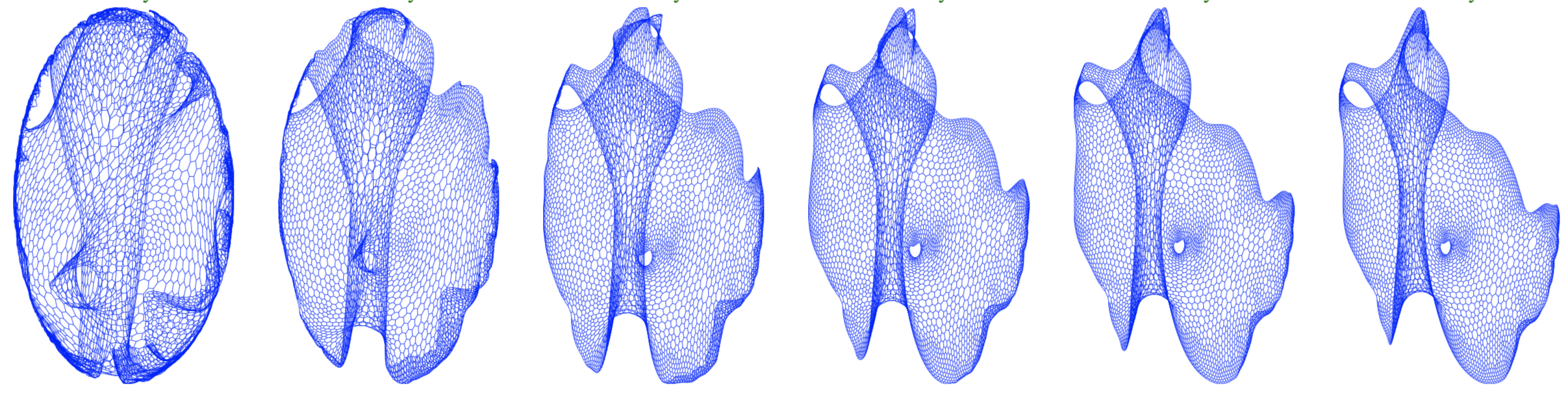}}                                                                                                                \\ \hline
    
            \textbf{BS=64} & \multicolumn{6}{|c|}{\includegraphics[height=2.2cm,width=0.8\linewidth]{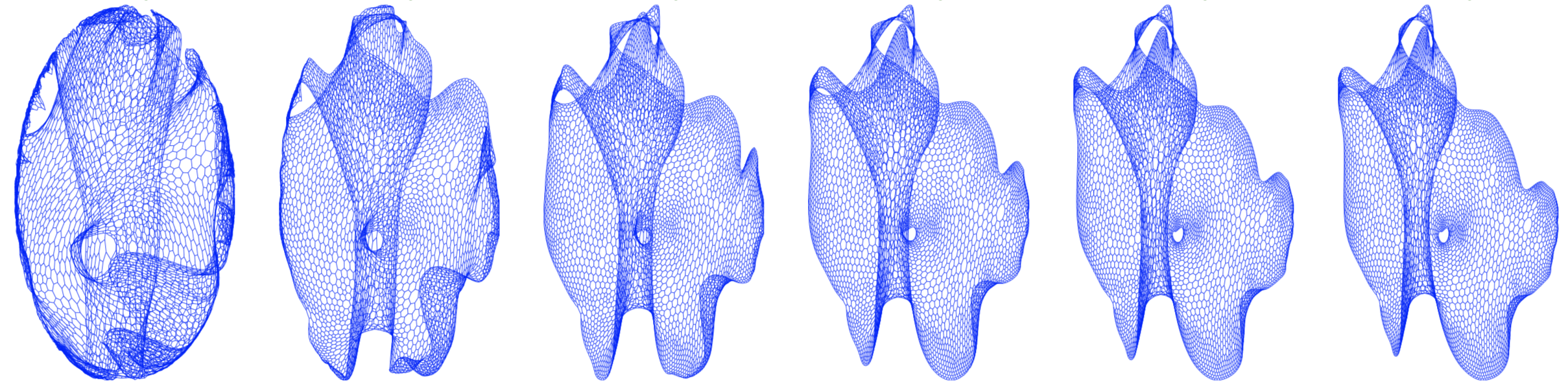}}                            \\ \hline
            
            \textbf{BS=256} & \multicolumn{6}{|c|}{\includegraphics[height=2.2cm,width=0.8\linewidth]{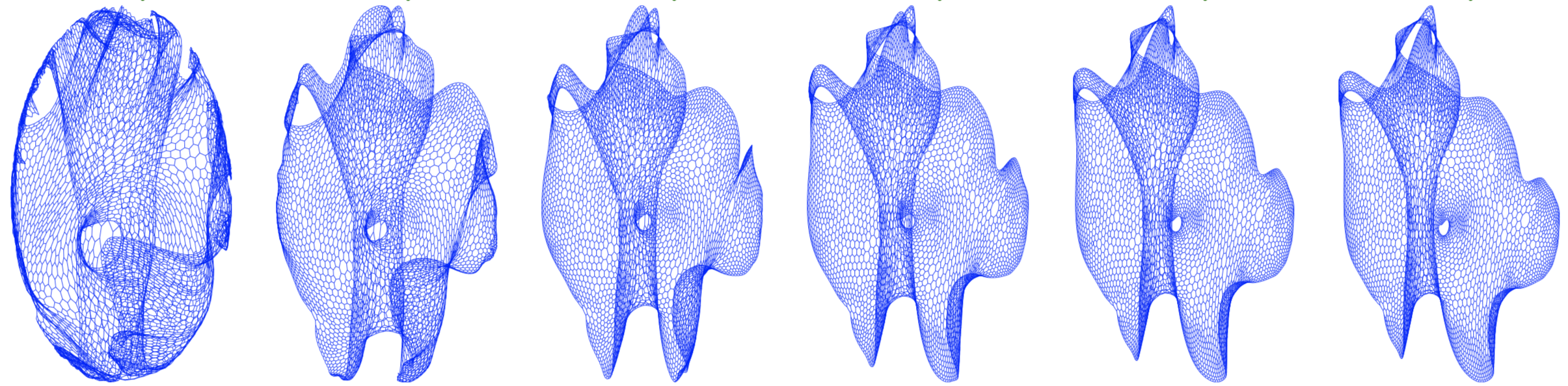}}                              \\ \hline
    \textbf{BS=1024} & \multicolumn{6}{|c|}{\includegraphics[height=2.2cm,width=0.8\linewidth]{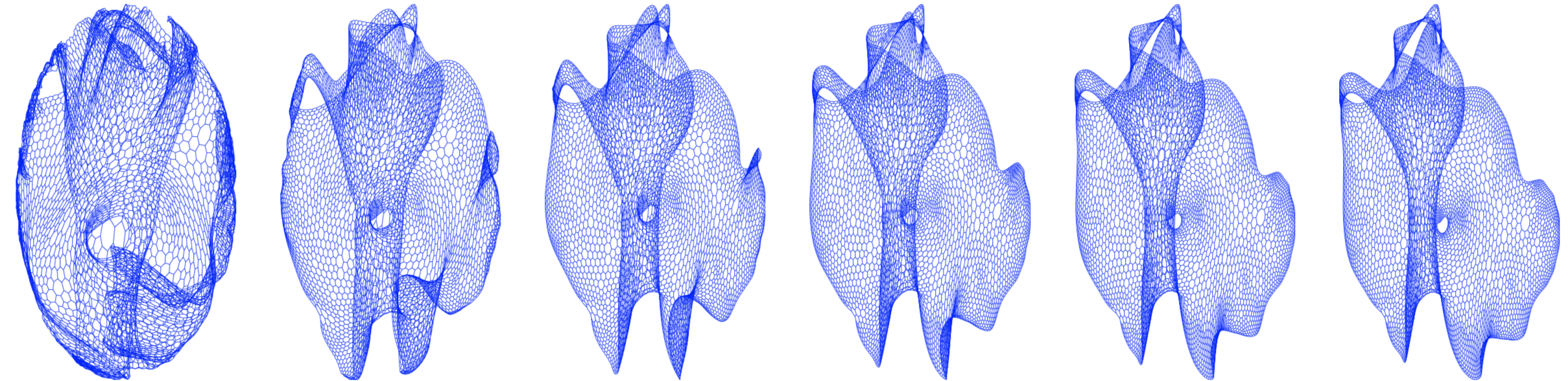}}                              \\ \hline
    \textbf{BS=2048} & \multicolumn{6}{|c|}{\includegraphics[height=2.2cm,width=0.8\linewidth]{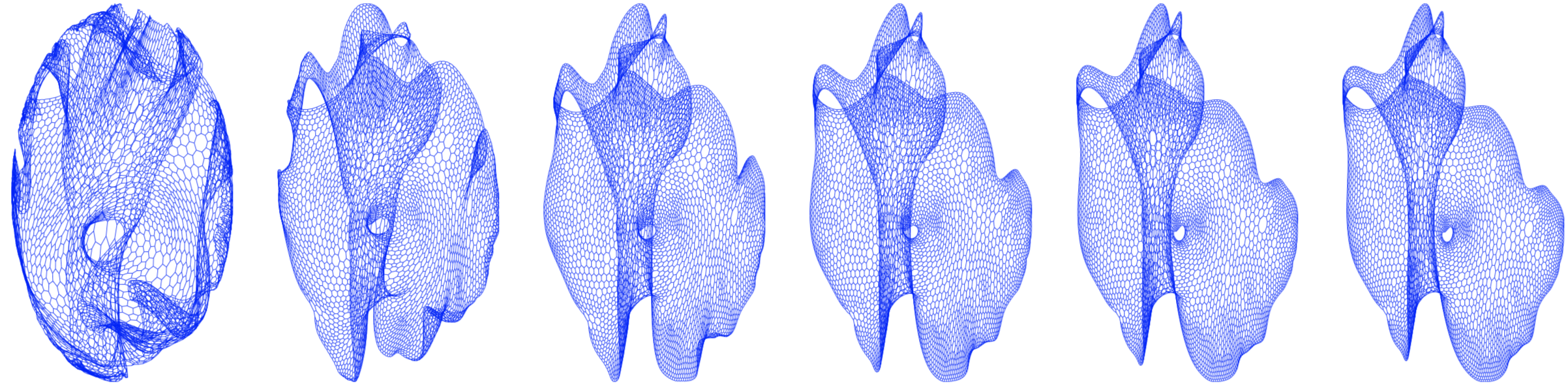}}                              \\ \hline
            
\end{tabular}
\end{table}

\section{How to run other tools}

We have used Gephi's toolkit to run ForceAtlas2 and OpenOrd. This integrated tool is available in our repository with proper documentation. Please check out our repository for more details. We have also used OpenOrd tool from authors repository which is written in C++ language using MPI.

\noindent{}\noindent{}\noindent{}To run Gephi's ForceAtlas2, we use commands like following:

\begin{markdown}
```
java -jar othertools/GephiLayouts-1.0.jar forceatlas2 -i 3elt_dual.gml -o 3elt_dual.forceatlas2.gml -threads 8 -maxiters 600
```
\end{markdown}
\newline
\noindent{}\noindent{}\noindent{}To run Gephi's OpenOrd, we use commands like following:

\begin{markdown}
```
java -jar othertools/GephiLayouts-1.0.jar openord -i 3elt_dual.gml -o 3elt_dual.forceatlas2.gml -threads 8 -maxiters 600
```
\end{markdown}
\newline
\noindent{}\noindent{}\noindent{}To run MPI version of OpenOrd, we use commands like following:

\begin{markdown}
```
mpirun --oversubscribe -n 48 ./bin/layout ./input/3elt_dual
```
\end{markdown}
\newline
\noindent{}\noindent{}\noindent{}To run tsNET, we use commands like following:

\begin{markdown}
```
python tsnet.py 3elt_dual.vna --output 3elt_dual.out.vna --perplexity 800 --learning_rate 6000
```
\end{markdown}

\begin{table*}[t]
\caption{Running time of BatchLayout of $O(n^2)$ version for different graphs. We set number of iterations to 500 and enable \emph{-ffast-math -mavx512f -mavx512dq} flags which are available in our Skylake server.}

\centering
\begin{tabular}{|c|c|c|c|}
\hline
\textbf{Graph} & \textbf{Runtime(sec.)} & \textbf{Graph} & \textbf{Runtime(sec.)} \\ \hline

Powergrid &	0.86	 &	add32 &	0.88	\\ \hline
ba\_network  &	1.27 &	3elt\_dual & 2.38 \\ \hline	PGPgiantcompo &	3.27 & pkustk02 &	3.34 \\ \hline
fe\_4elt2 &	3.43	&		bodyy6 &	9.49 \\ \hline	pkustk01 &		12.27 &	OPF\_6000 &		22.37 \\ \hline finance256  & 34.5 &		finan512 & 131.73\\ \hline
\end{tabular}
\label{tab:measures_batchlayout_time}
\end{table*}

\begin{table*}
\caption{Running time of tsNET for different graphs. tsNET failed to run for finance256 and other graphs which have more vertices than finance256.}

\centering
\begin{tabular}{|c|c|c|c|}
\hline
\textbf{Graph} & \textbf{Runtime(sec.)} & \textbf{Graph} & \textbf{Runtime(sec.)} \\ \hline

Powergrid &	1465.63	 &	add32 &	712.68	\\ \hline
ba\_network  &	2208.14 &	3elt\_dual & 3336.54 \\ \hline	PGPgiantcompo &	7546.79 & pkustk02 &	4195.17 \\ \hline
fe\_4elt2 &	5707.4	&		bodyy6 &	26923.84 \\ \hline	pkustk01 &		26953 &	OPF\_6000 &		60557.65 \\ \hline finance256  & - &		finan512 & -\\ \hline
\end{tabular}
\label{tab:measures_tsnet_time}
\end{table*}

\begin{table*}
\caption{Comparision of Edge uniformity (EU) measures among random initialized version of \toolname{}. For this measure, lower value means better result.}

\centering
\begin{tabular}{|c|c|c|c|c|c|}
\hline
\multirow{1}{*}{\textbf{Graph}} & \multicolumn{2}{c|}{\textbf{EU}}  &  \multirow{1}{*}{\textbf{Graph}}
& \multicolumn{2}{c|}{\textbf{EU}}            \\ \cline{2-3} \cline{5-6}
                                & BLR       & BLRBH & & BLR & BLRBH  \\ \hline

Powergrid &		0.784 &		0.529 & plustk02	& 	0.844 &		0.699\\ \hline
add32 &	1.354 &		1.029 &	 fe\_4elt2 &		0.404 &		0.51\\ \hline
ba\_network	 &	1.016 &		0.566 & bodyy6 &	0.741 &		0.806 \\ \hline
3elt\_dual &		0.361 &		0.443 &	pkustk01 &	0.76 &		0.659\\ \hline
PGPgiantcompo &	0.812 &		0.704 & - &		- &	-\\ \hline

\end{tabular}
\label{tab:measures_random_init0}
\end{table*}

\begin{table*}
\caption{Comparision of Stress (ST) and Neighborhood preservation (NP) measures among random initialized version of \toolname{}. For ST, lower value means better result and for NP, higher value represents better result. }
\centering
\begin{tabular}{|c|c|c|c|c|}
\hline
\multirow{1}{*}{\textbf{Graph}} & \multicolumn{2}{c|}{\textbf{ST}}                            & \multicolumn{2}{c|}{\textbf{NP}}            \\ \cline{2-5} 
                                & BLR       & BLRBH & BLR & BLRBH  \\ \hline

Powergrid	&	1.7E+6	&	2.2E+6	&	0.336 &		0.208\\ \hline
add32	&	2.0E+6	&	2.7E+6	&	0.437	&	0.229\\ \hline
ba\_network	&	2.9E+6	&	2.8E+6	&	0.346	&	0.283\\ \hline
3elt\_dual	&	1.0E+7 &	1.3E+7	&	0.168	&	0.112\\ \hline
PGPgiantcompo	&	8.1E+6 &		1.7E+7	&	0.182	&	0.109\\ \hline
plustk02	&	2.7E+7	&	4.7E+7	&	0.412	&	0.211\\ \hline
fe\_4elt2	&	1.6E+7	&	2.1E+7	&	0.268	&	0.165\\ \hline
bodyy6	&	3.9E+7	&	4.2E+7	&	0.171	&	0.177\\ \hline
pkustk01	&	4.6E+6	&	4.2E+6	&	0.626	&	0.599\\ \hline

\end{tabular}
\label{tab:measures_random_init}
\end{table*}

\begin{figure}
    \centering
    \includegraphics[width=\linewidth]{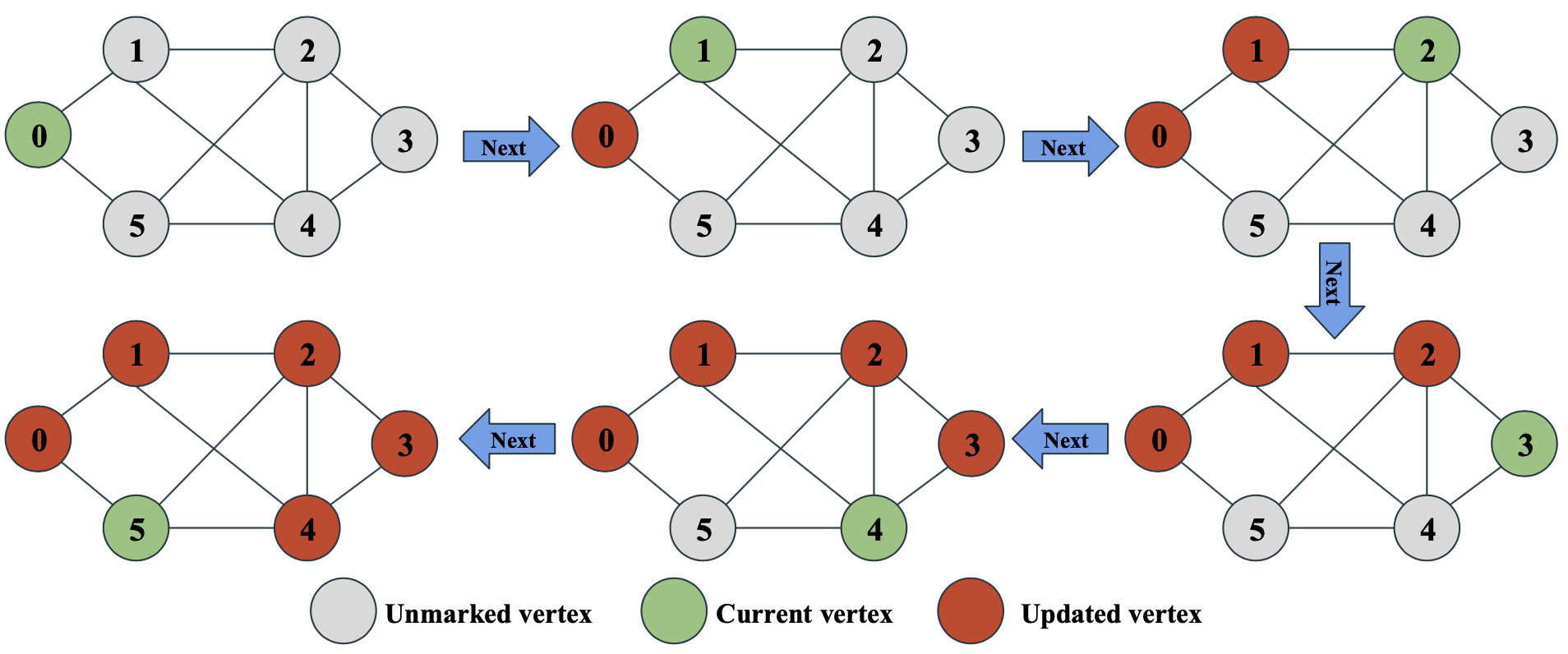}
    \caption{Example of force calculations in sequential approach. Gray colored vertex represents unmarked vertex whose updated position has neither been calculated nor is being calculated. Green colored vertex represents current vertex whose forces are being calculated with respect to all other vertices. Red colored vertex represents those cases whose forces have already been updated/calculated.}
    \label{fig:sgdfig}
\end{figure}

\begin{figure*}
    \centering
    \includegraphics[width=0.32\linewidth]{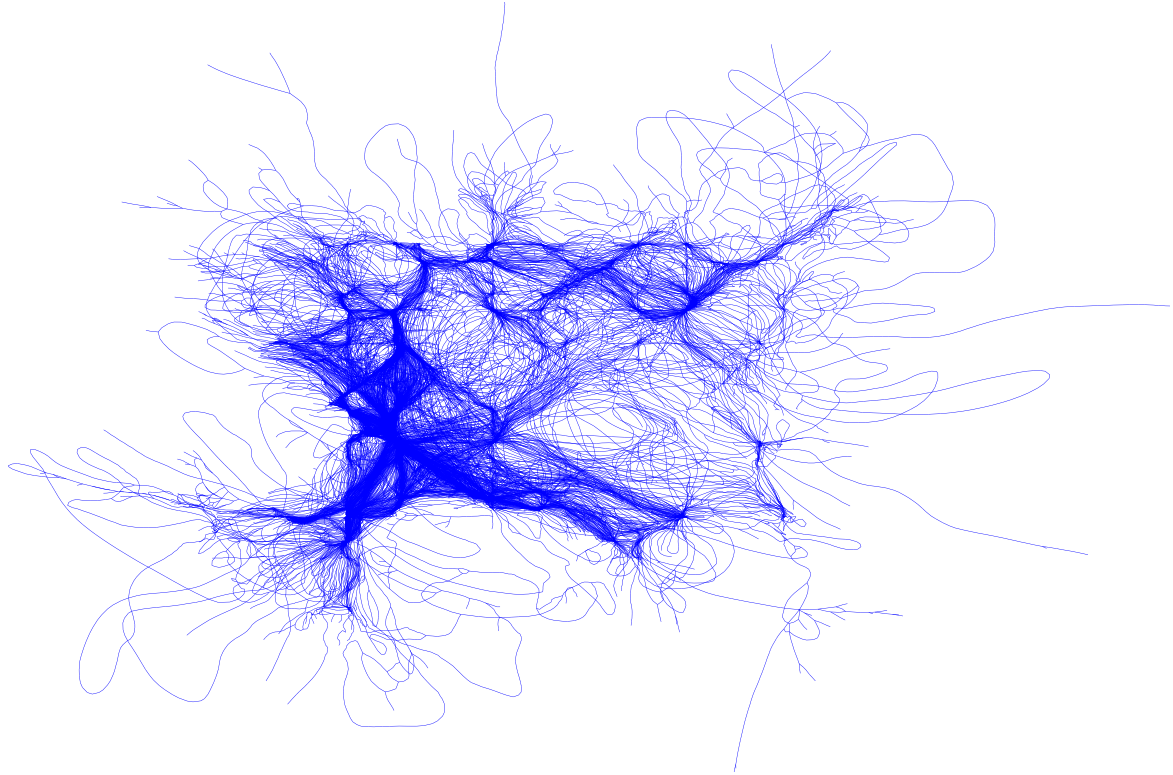}
    \includegraphics[width=0.32\linewidth]{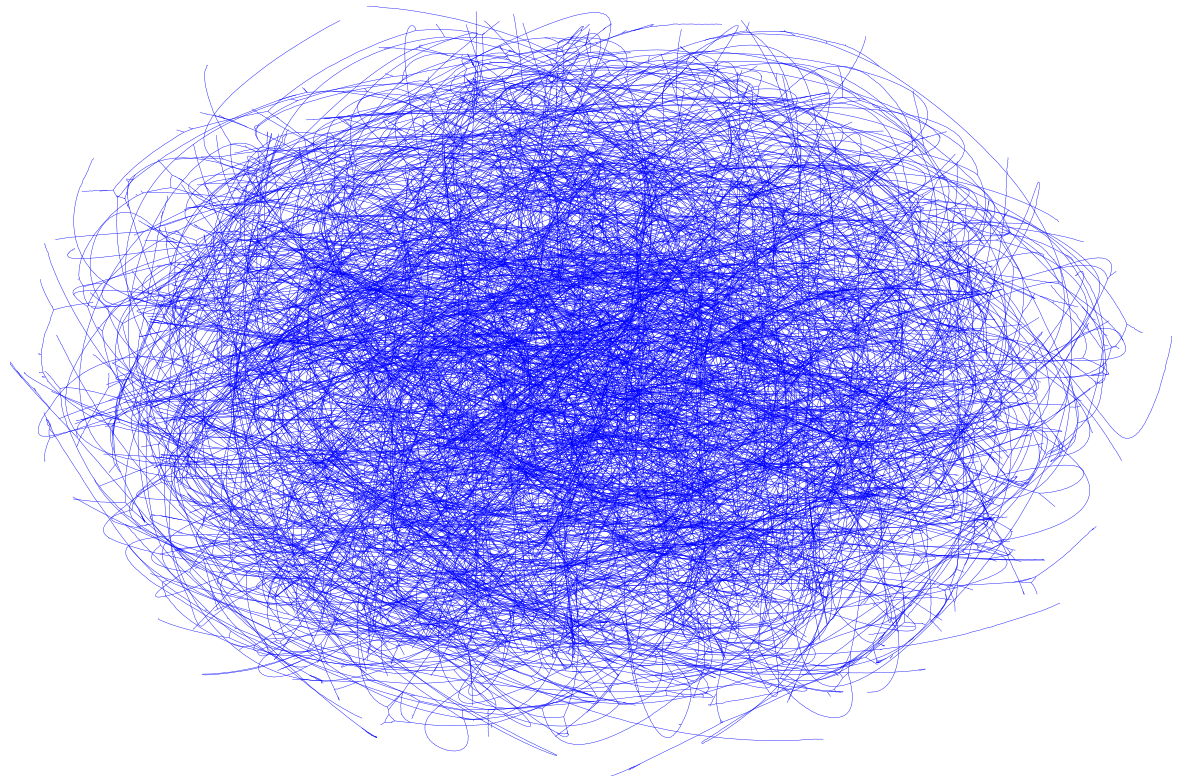}
    \includegraphics[width=0.32\linewidth]{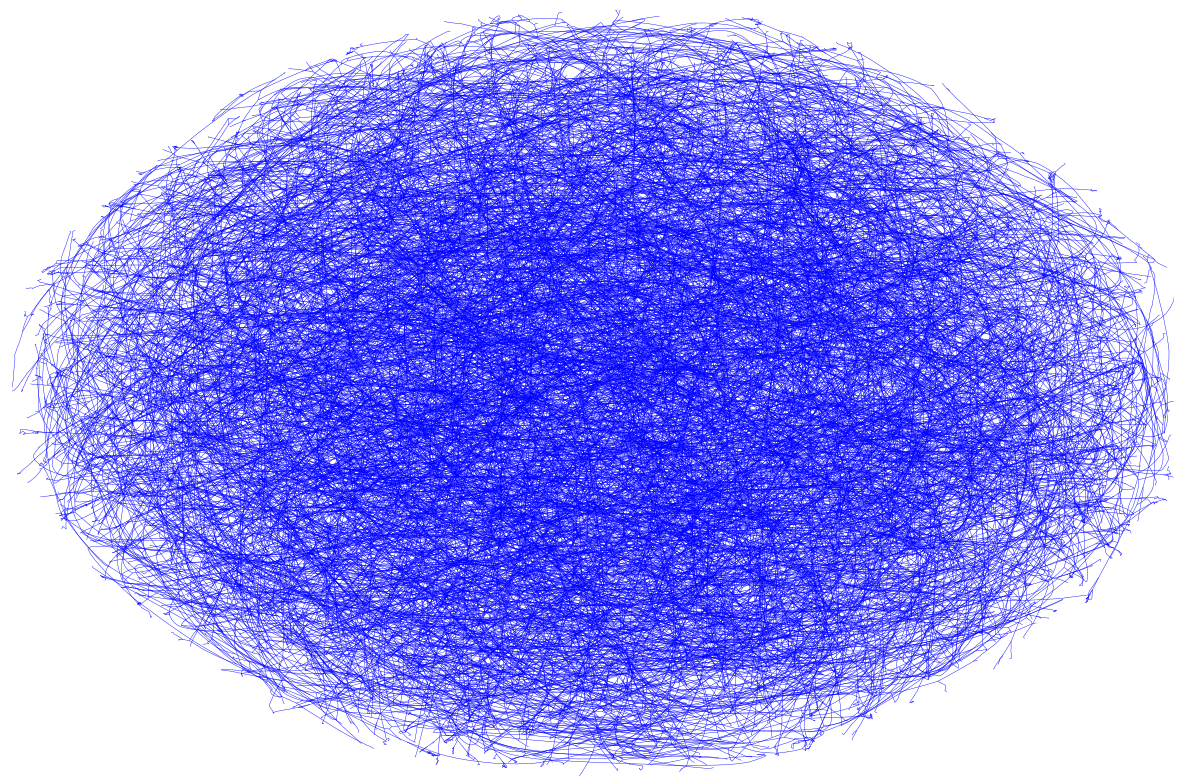}
    \caption{Layouts of lxOSM which has 114,599 vertices and 239,332 edges. BatchLayoutBH (left plot), ForceAtlas2BH (middle plot) and OpenOrd (right plot) took 24.75 seconds, 102.0 seconds and 50.03 seconds, respectively for 750 iterations using 48 threads. \toolnameBH{} is capable of generating readable layout within very short time than other two tools.}
    \label{fig:biggraphs}
\end{figure*}

\begin{figure*}
    \centering
    \includegraphics[width=0.48\linewidth]{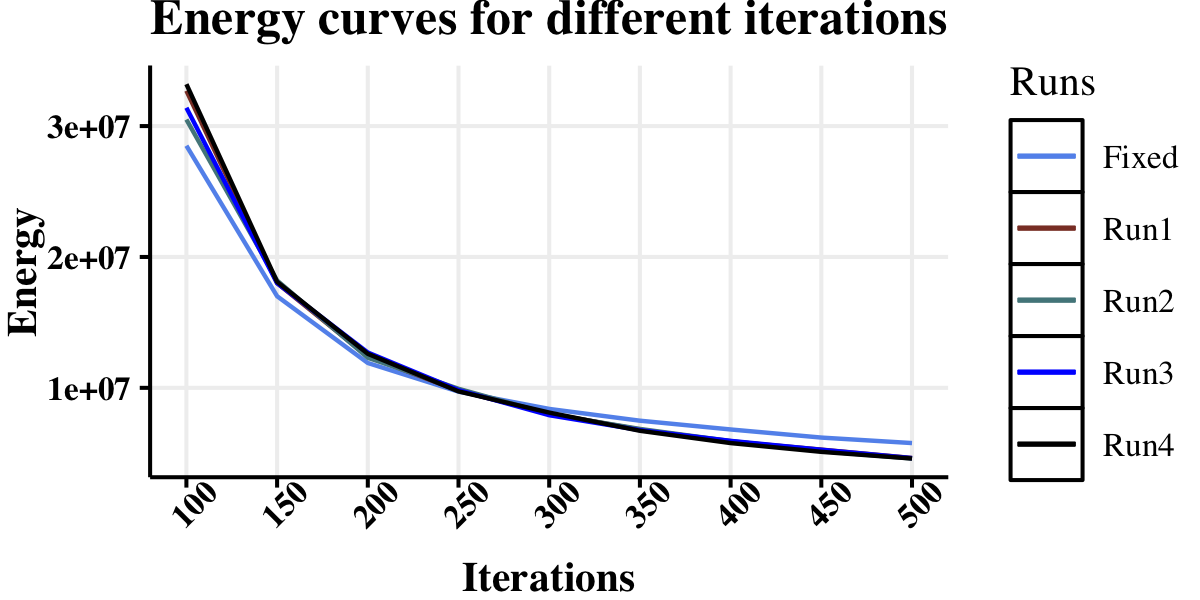}
     \includegraphics[width=0.48\linewidth]{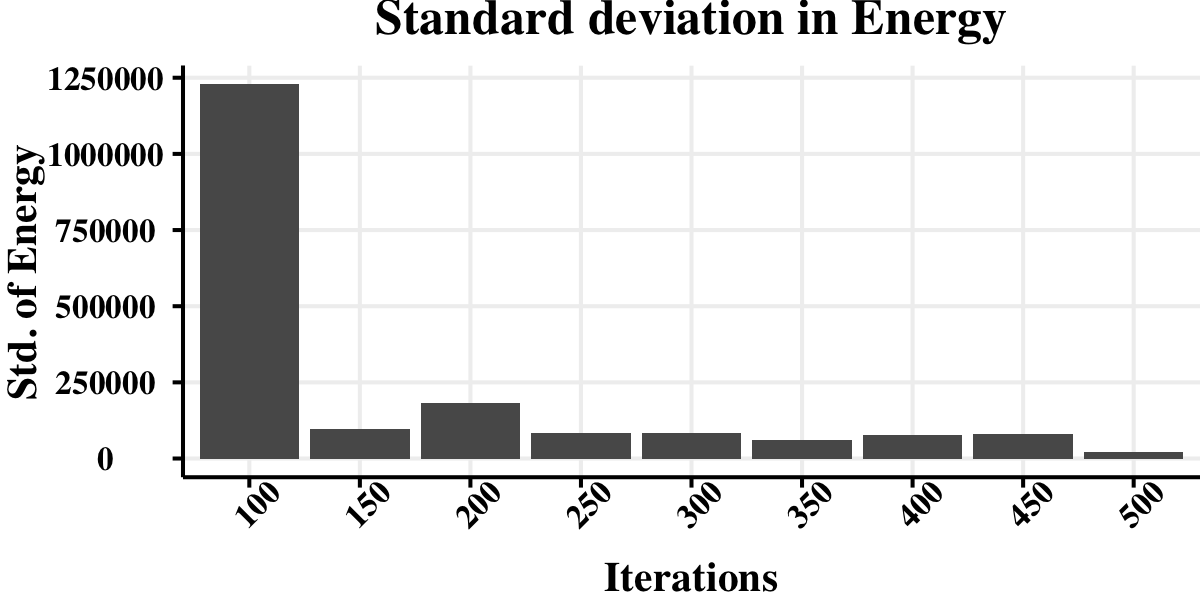}
    \caption{Energy curves of random batch selection for different runs with same hyper-parameters (3elt\_dual graph) (left figure). There are differences but due to high value of energy, it is not clearly visible and we demonstrate this difference in right figure.}
    \label{fig:origruns}
\end{figure*}

\begin{figure}
    \centering
    \includegraphics[width=0.7\linewidth]{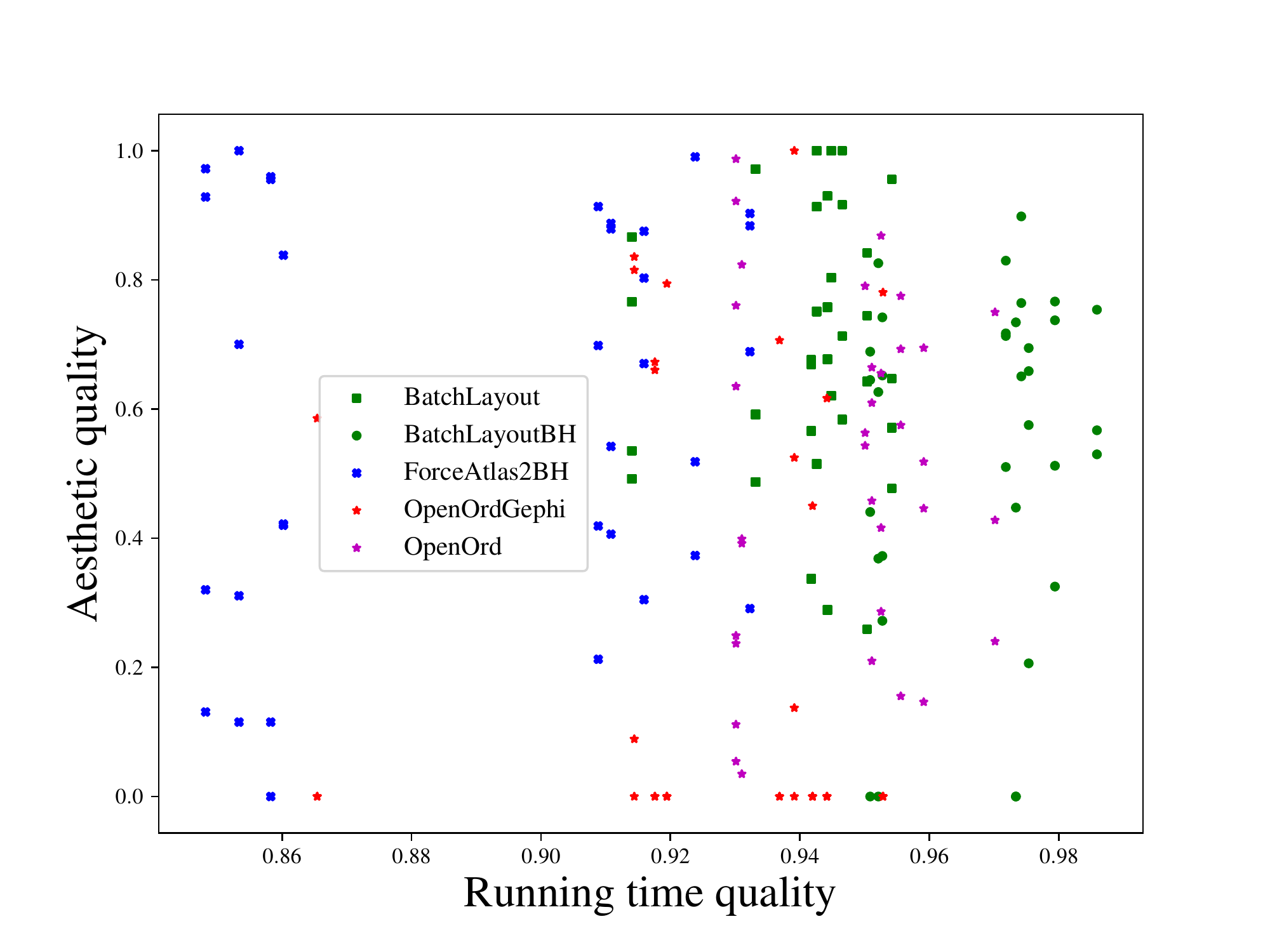}
    \caption{Summary of running time and aesthetic quality of all tools for 500 iterations discussed in the manuscript, Tables II, III and IV. We normalize running time by dividing the maximum running time among all tools for a corresponding graph and subtracting from 1 to feel like higher number is better for quality. We similarly compute other normalized aesthetic measure to set within range 0 to 1 where higher value means better quality. We put normalized running time quality in x-axis and normalized aesthetic measures in y-axis for each corresponding graph. ForceAtlas2 is not shown due to its lower normalized value for running time which is always 0. Obviously, the best tool will touch the right-top corner i.e., better running time as well as better quality. BatchLayout and BatchLayoutBH are shown in same color to distinguish from other tools. We clearly see that BatchLayout and BatchLayoutBH attain better running time and quality.}
    \label{fig:summary500iterations}
\end{figure}

\begin{algorithm}
\caption{constructBHT}
\hspace*{\algorithmicindent} \textbf{Input:} $c$, $D$
\begin{algorithmic}[1]
\State $p_1 \leftarrow$ consecutive $c_i$'s in left top cell
\State $p_2 \leftarrow$ consecutive $c_i$'s in right top cell
\State $p_3 \leftarrow$ consecutive $c_i$'s in left bottom cell
\State $p_4 \leftarrow$ consecutive $c_i$'s in right bottom cell
\State compute diameter $D = \frac{D}{2}$
\State $t_1 \leftarrow$ constructBHT($p_1$, $D$) in parallel
\State $t_2 \leftarrow$ constructBHT($p_2$, $D$) in parallel
\State $t_3 \leftarrow$ constructBHT($p_3$, $D$) in parallel
\State $t_4 \leftarrow$ constructBHT($p_4$, $D$) in parallel
\State check coordinates if any of $t_i$'s is leaf node
\State compute centroid $s$ using $t_1$,$t_2$,$t_3$,and $t_4$
\newline
\Return create tree \emph{node} using $t_1$,$t_2$,$t_3$,$t_4$, $s$, and $D$
\end{algorithmic}
\label{algo:constructbht}
\end{algorithm}

\begin{algorithm}
\caption{calcRepulsiveForce}
\hspace*{\algorithmicindent} \textbf{Input:} $r$, $i$, $f$
\begin{algorithmic}[1]
\State $d = \parallel r_c - i\parallel$ \Comment{Euclidean distance}
\If{$\theta > \frac{D}{d}$} \Comment{MAC condition}
\State $f += f_a(r_c,i)\times \frac{r_c-i}{d}$ \Comment{$r_c$ indicates coordinates of $r$}
\Else 
\For{$child$ of $r$}
\If{$child$ has descendent}
\State calcRepulsiveForce($child$, $i$, $f$)
\Else
\State $f += f_r(child_c,i)\times \frac{child_c-i}{d}$ \Comment{$child_c$ indicates coordinates of $child$}
\EndIf
\EndFor
\EndIf
\end{algorithmic}
\label{algo:queryserve}
\end{algorithm}

\begin{algorithm}
\caption{BatchLayoutBH}
\hspace*{\algorithmicindent} \textbf{Input:} G(c, V, E)
\begin{algorithmic}[1]
\State $Step = 1.0$ \Comment{initial step length}
\State $Loop = 0$
\While {$ Loop < MaxIteration$}
\State $Energy = 0$
\State re-scale $c$ in bounding box
\State sort $c$ in parallel based on morton order
\State $D = max(max(c_x)-min(c_x),max(c_y)-min(c_y))$
\State $BHT = constructBHT(c, D)$ \Comment{call Algorithm \ref{algo:constructbht}}
\ParFor{$i \in V$}
			\State $ct_{i} = 0$
\EndParFor
\For{$b \leftarrow$ 0 to $\frac{n}{BS}-1$}
    \ParFor{$i \leftarrow b * BS$ to $(b + 1) * BS-1$ \textbf{by} $bx$}
        \State $f = (0,0)$
        \For{$j\in G.colids[i]$}
                    \State $f\;+= f_a(i,j)\times \frac{c_j - c_i}{||c_i - c_j||}$
        \EndFor 
        \State $calcRepulsiveForce(BHT, i, rf = (0,0))$ \Comment{call Algorithm \ref{algo:queryserve}}
        \State $f = f - rf$;
        \State $ct_{i+x} += f$
    \EndParFor
    \ParFor{$i \leftarrow b * BS $ to $(b + 1) * BS - 1$}
        \State $c_i\; += Step \times \frac{ct_i}{||ct_i||}$
        \State $Energy\; += ||ct_i||^2$
    \EndParFor
\EndFor
\EndWhile
\newline
\Return the final layout $c$
\end{algorithmic}
\label{algo:cbBatchLayout}
\end{algorithm}

\begin{figure}
    \centering
    \includegraphics[width=0.3\linewidth]{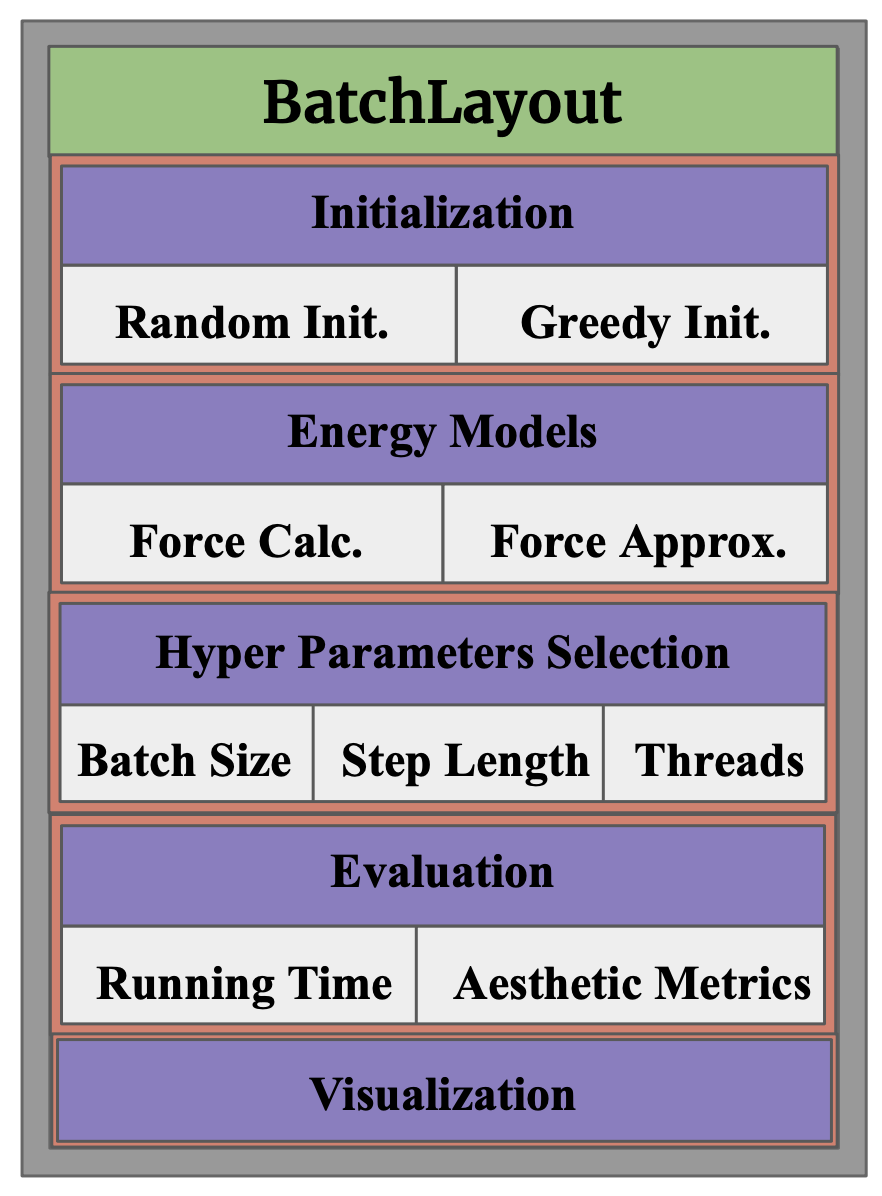}
    \caption{BatchLayout system overview. In BatchLayout, we demonstrate two initialization techniques, several energy models and repulsive force approximation techniques, flexibility of choosing batch size and number of threads, evaluation based on run time and aesthetic metrics, and visualization.}
    \label{fig:blsystem}
\end{figure}

\begin{table*}[t]
\caption{Running time of $FM^3$ of OGDF library for different graphs. We run default option of $FM^3$ by setting number of iterations to 500.}

\centering
\begin{tabular}{|c|c|c|c|}
\hline
\textbf{Graph} & \textbf{Runtime(sec.)} & \textbf{Graph} & \textbf{Runtime(sec.)} \\ \hline

Powergrid &	2.21	 &	add32 &	1.34	\\ \hline
ba\_network  &	2.45 &	3elt\_dual & 3.77 \\ \hline	PGPgiantcompo &	5.90 & pkustk02 &	5.27 \\ \hline
fe\_4elt2 &	4.13	&		bodyy6 &	7.75 \\ \hline	pkustk01 &		9.82 &	OPF\_6000 &		11.68 \\ \hline finance256  & 13.33 &		finan512 & 27.89\\ \hline
comYout.  & 897.17 &		Flan\_1565 & 1105.33\\ \hline
\end{tabular}
\label{tab:measures_fm3_time}
\end{table*}

\begin{table*}[]
\caption{Number of edge crossing in the layouts after running each tool for 500 iterations.}
\centering
\begin{tabular}{|c|c|c|c|c|c|}
\hline
\textbf{Graph} & \textbf{BL} & \textbf{BLBH} & \textbf{FA2} & \textbf{FA2BH} & \textbf{OO} \\ \hline
Powergrid      & 3892        & 8030          & 3300         & 3308           & 6631        \\ \hline
add32          & 5745        & 22466         & 1325         & 1608           & 17145       \\ \hline
ba\_network    & 526         & 7518          & 273          & 334            & 588         \\ \hline
3elt\_dual     & 6208        & 12678         & 13992        & 8385           & 24990       \\ \hline
PGPgiant.      & 909233      & 2096136       & 626802       & 632239         & 722747      \\ \hline
fe\_4elt2      & 51472       & 252197        & 67624        & 73995          & 133574      \\ \hline
bodyy6         & 208077      & 363718        & 153062       & 181133         & 475689      \\ \hline
\end{tabular}
\end{table*}